%% file: main.tex
\documentclass[conference]{IEEEtran}

\IEEEoverridecommandlockouts  

\usepackage[utf8]{inputenc}

\pdfoutput=1

\usepackage{cite}
\usepackage{multicol}
\usepackage[bookmarks=true]{hyperref}

\usepackage{mathrsfs} 
\usepackage{amsmath,amssymb,amsthm,amsfonts,graphicx,float}
\usepackage{mathtools}
\usepackage{color} 
\usepackage{tabularx}

\usepackage{algorithm}
\usepackage{algpseudocode}

\usepackage{tikz,pgfplots}
\pgfplotsset{compat=1.8}
\usepackage{pgfplotstable}
\usepgfplotslibrary{groupplots}
\usepackage{times}

\newcommand{\myqedblock}{\hfill$\square$}

\newtheorem{theorem}{Theorem}
\newtheorem{assumption}{Assumption}
\newtheorem{sideinfo}{Side information}

\newtheorem{corollary}{Corollary}

\newtheorem{definition}{Definition}
\newtheorem{example}{Example}

\newtheorem{lemma}{Lemma}

\newtheorem{problem}{Problem}

\newtheorem{remark}{Remark}

\newcommand{\reachalgo}{\texttt{DaTaReach}}
\newcommand{\controlalgo}{\texttt{DaTaControl}}

\newcommand{\myextended}{Appendix~\ref{appendix-A}}

\usepackage{placeins}

\begin{document}

\title{\LARGE \bf On-The-Fly Control of Unknown Smooth Systems from Limited Data
    \thanks{This material is based on work supported by Air Force Office of Scientific Research and DARPA Assured Autonomy Program (FA9550-19-1-0005), and National Aeronautics and Space Administration (80NSSC19K0209).\newline 
    \indent F. Djeumou, A. Vinod, and U. Topcu are with the Department of Electrical and Computer Engineering, Oden Institute for Computational Engineering and Sciences, and the Department of Aerospace Engineering and Engineering Mechanics at the University of Texas at Austin, Austin, TX, USA. Email: \texttt{\{fdjeumou, utopcu\}@utexas.edu}, \texttt{aby.vinod@gmail.com}.\newline
    \indent E. Goubault and S. Putot are with the LIX, CNRS, \'Ecole polytechnique, Institut Polytechnique de Paris, France. Email: \texttt{\{goubault, putot\}@lix.polytechnique.fr}.}}
\author{Franck Djeumou, Abraham P. Vinod, Eric Goubault, Sylvie Putot, and Ufuk Topcu}
\maketitle

\begin{abstract}
    We investigate the problem of data-driven, on-the-fly control of systems with unknown nonlinear dynamics where data from only a single finite-horizon trajectory and possibly side information on the dynamics are available. Such side information may include knowledge of the regularity of the dynamics, monotonicity of the states, or decoupling in the dynamics between the states. Specifically, we develop two algorithms, \reachalgo{} and \controlalgo{}, to over-approximate the reachable set and design control signals for the system on the fly. \reachalgo{} constructs a differential inclusion that contains the unknown vector field. Then, it computes an over-approximation of the reachable set based on interval Taylor-based methods applied to systems with dynamics described as differential inclusions. \controlalgo{} enables convex-optimization-based, near-optimal control using the computed over-approximation and the receding-horizon control framework. We provide a bound on its suboptimality and show that more data and side information enable \controlalgo{} to achieve tighter suboptimality bounds. Finally, we demonstrate the efficacy of \controlalgo{} over existing approaches on the problems of controlling a unicycle and quadrotor systems.
\end{abstract}

\section{Introduction}

Consider a scenario in which abrupt changes in the dynamics of a system occurs. The changes in the dynamics are such that the a priori known model cannot be used, and there is a need to learn the new dynamics on the fly. In such a scenario, the system needs to retain a certain degree of control with data from only its current trajectory. This paper considers the problem of data-driven, on-the-fly control of systems with unknown dynamics under severely limited data.

\begin{figure}[t]
    \centering
    \includegraphics[width=1\linewidth]{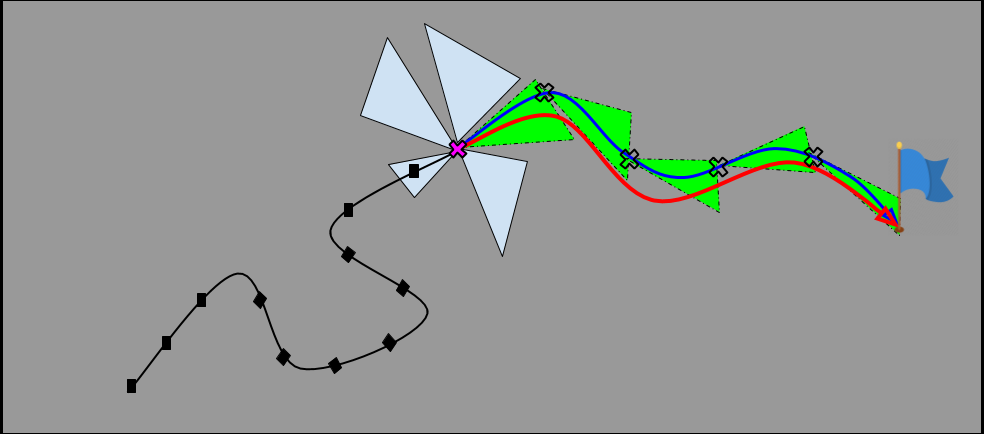} 
    \label{fig:prob_st}
    \begin{tikzpicture}[remember picture, overlay, text
        width=9em]
        \small
        \node [text centered, xshift=13em, yshift=10em]
            (Goal) {Goal};
        \node [text centered, xshift=-9.2em,
            yshift=11.5em] (Overapproximation)
            {Over-approximations for various controls for all $t \in [t_N,t_{N+1}]$};
        \node [text centered, xshift=-9em, yshift=8.5em,
            text width=7em] (System) {State at $t_i = t_N$};
        \node [text width=18em, text centered,
            xshift=-5em, yshift=2em] (Trajectory)
            {Sampled trajectory for $t < t_N$};
        \node [text centered, xshift=7em, yshift=12.5em, text width=10.5em] 
            (nearoptimal) {Over-approximations for near-optimal controls};
        \node [text centered, xshift=8em, yshift=3em] 
            (Control) {Near-optimal control synthesis for
            $t \geq t_N$};
        \node [text centered, xshift=9em, yshift=5.5em] 
            (Optimal) {Optimal trajectory};
        \draw [-latex]  (System.east) -- 
            ++ (4.15em, 0.9em);
        \draw [-latex]  (Overapproximation.east) -- 
            ++ (2em, -1em);
        \draw [-latex]  (Overapproximation.east) -- 
            ++ (4em, -0.10);
        \draw [-latex]  ([xshift=0em,
            yshift=0.2em]nearoptimal.south) -- ++ (1.7em, -1.8em);
        \draw [-latex]  ([xshift=0em,
            yshift=0.2em]nearoptimal.south) -- ++ (-1.7em, -2.4em);
        \draw [-latex]  ([xshift=0.7em,
            yshift=0.7em]Control.west) -- ++ (-1.2em, 7.0em);
        \draw [-latex]  ([xshift=0.7em,
            yshift=0.7em]Optimal.west) -- ++ (-0.6em, 2em);
    \end{tikzpicture}
    \vspace*{-1em}
    \caption{We use limited data and side information for on-the-fly control of systems with unknown dynamics. At each sampling time, we compute an over-approximation of the set of states the system may reach to describe the uncertainty in its unknown trajectory and construct one-step optimal controllers via convex optimization.}
    \vspace*{-4mm}
\end{figure}
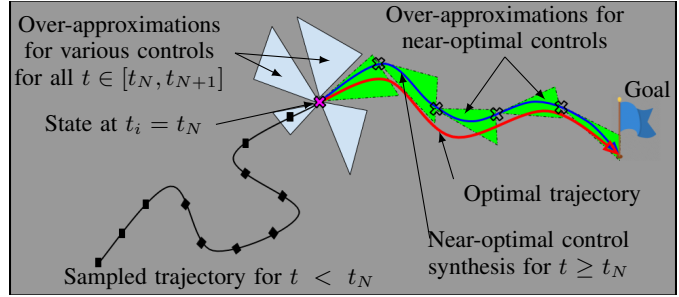

We develop data-driven algorithms, \reachalgo{} and \controlalgo{}, to over-approximate the reachable set and control systems with unknown dynamics. Reachability analysis provides a framework to analyze the set of states reached by a dynamical system. We use such a framework as the baseline for the on-the-fly control, as it enables robust control to perturbations on the data. When the underlying dynamics are known, Hamilton-Jacobi-based~\cite{mitchell2005time} or interval Taylor-based~\cite{berz1998verified,nedialkov1999validated,goubault2019inner} methods can approximate the reachable set. However, limited work has been reported for the case where the underlying dynamics are not known a priori, and even fewer work considers settings with severely limited data.

\reachalgo{} and \controlalgo{} can work with data from a single finite-horizon trajectory of the system and take advantage of various forms of side information on the dynamics. The algorithms consider trajectories containing finite samples of the states, the derivatives of the states, and the control signals applied. Furthermore, if available, they can make use of side information such as knowledge of the regularity of the dynamics, bounds on the vector field, monotonicity of the vector field, decoupling in the dynamics among the states of the system, or knowledge of parts of the dynamics. Such side information can be extracted from data, responses of the system to inputs, or known elementary laws of physics.

\reachalgo{} provides closed-form expressions for over-approximations of the reachable set of unknown dynamical systems. It first utilizes the available data and the given side information to construct differential inclusions that contain the vector field of the unknown dynamics. Unlike existing work~\cite{rezaTAC2020} that provides state-independent differential inclusions, the constructed differential inclusions account for dependencies on the states and control signals. Then, it provides closed-form expressions for over-approximations of the reachable set associated with these differential inclusions. To obtain such over-approximations, it utilizes interval Taylor-based methods~\cite{berz1998verified,nedialkov1999validated,goubault2019inner} for known dynamics to compute over-approximations of the reachable set of dynamics described by these differential inclusions. The closed-form expressions can incorporate the side information, and more data and side information provide tighter over-approximations.

The closed-form expressions enable convex-optimization-based, near-optimal control of unknown dynamical systems based on the one-step receding-horizon control framework~\cite{mayne2000constrained}. Specifically, we seek to sequentially minimize a given one-step cost function in a discrete-time setting. The one-step cost function, which encodes the desired behavior of the system, has to be optimized in a black-box manner since it is typically a function of the \emph{unknown next state}, the known current state, and the current input. \controlalgo{} computes \emph{approximate solutions} to such an optimization problem via convex optimization relaxations. We obtain such relaxations by replacing the unknown next state with a control-affine linearization of the corresponding over-approximation of the reachable set. Furthermore, we provide a bound on the suboptimality of the approximate solutions and show that such a bound becomes tighter when more data and side information are available.

\noindent {\textbf{Related work.}
For the problem of data-driven control, researchers have proposed several approaches that combine system identification with model predictive control~\cite{korda2018linear,kaiser2018sparse,proctor2016dynamic,ornik2019myopic,vinod2020convexified}. In~\cite{korda2018linear}, the authors use Koopman theory to lift the unknown nonlinear dynamics to a higher-dimensional space where they perform linear system identification. \texttt{SINDYc}~\cite{kaiser2018sparse} utilizes a sparse regression over a library of nonlinear functions for nonlinear system identification. \texttt{DMDc}~\cite{proctor2016dynamic} uses the spectral properties of the collected data to obtain approximate linear models. Thus, \texttt{DMDc} can perform poorly due to the restriction to linear dynamics. Myopic control~\cite{ornik2019myopic} uses a finite sequence of perturbations to learn a local linear model, which is then used to optimize a model-driven goodness function that encodes desirable behaviors. \texttt{DMDc}, \texttt{SINDYc}, and approaches based on the Koopman theory require significantly more data than the proposed approach. Myopic control cannot incorporate any of the side information mentioned above. Similarly, in their current forms, we believe that \texttt{DMDc}, \texttt{SINDYc}, and approaches based on the Koopman theory cannot incorporate the side information. Furthermore, the experiments show that \controlalgo{} performs significantly better than \texttt{SINDYc}.

Contextual optimization-based approaches tackle the data-driven control problem via surrogate optimization and skip the system identification step~\cite{vinod2020convexified,krause2011contextual}. These approaches minimize the one-step cost function in a black-box manner using the data. \texttt{C2Opt}~\cite{vinod2020convexified} exploits the structure in the given problem and utilizes side information (e.g., smoothness) and convex optimization to solve this problem. It overcomes the drawbacks of the Gaussian process-based approaches~\cite{krause2011contextual}, namely high computational costs, expensive hyperparameter tuning, and the inability to incorporate side information. Unlike \controlalgo{}, \texttt{C2Opt} considers limited forms of side information and relies on the knowledge of the gradient of the one-step cost, which may not be accessible. We demonstrate on a unicycle and quadrotor systems that \controlalgo{} is \emph{three orders of magnitude} computationally faster and less suboptimal than \texttt{C2Opt} and Gaussian process-based algorithms.

Recent work~\cite{devonport2020data,haesaert2017data,chakrabarty2018data} has considered the problem of data-driven estimation of the reachable sets of partially unknown dynamical systems. The approaches in these work rely on a system identification using either supervised learning algorithms~\cite{chakrabarty2018data} or Gaussian process-based algorithms~\cite{haesaert2017data, devonport2020data}. Such approaches are unable to take advantage of the side information, require significantly more data than \reachalgo{}, and provide only probabilistic guarantees of the correctness of the computed reachable sets while \reachalgo{} provides correct over-approximations at the expense of being conservative.

In Appendix~\ref{appendix-A}, we provide the proofs of all the technical results of the paper.

}

\section{Preliminaries and Problem Statement} \label{sec:prelim}

We denote an interval by $[a , b] = \{ x \in \mathbb{R} |  a \leq x \leq b \}$ for some $a, b\in \mathbb{R}$ such that $a \leq b$, the set $\{i,\hdots,j\}$ by $\mathbb{N}_{[i,j]}$ for $i,j \in \mathbb{N}$ with $i\leq j$ , the $2$-norm by $||\cdot||_2$, the $k^\mathrm{th}$ component of a vector $x$ and the $(k,j)$ component of a matrix $X$ by ${(x)}_k$ and ${(X)}_{k,j}$, respectively, the Lipschitz constant of $f : \mathcal{X} \to \mathbb{R}$ by $L_f = \sup \{ L \in \mathbb{R} \: \vert \: |f(x) - f(y)| \leq L \|x-y\|_2, x,y \in \mathcal{X}, x \neq y\}$ for $\mathcal{X} \subseteq \mathbb{R}^n$, and the Jacobian of a function $f$ by $\frac{\partial f}{\partial x}$. A function $f \in \mathscr{C}^k(\mathcal{X})$, referred as $f$ is $\mathcal{C}^{k}$, with $k\geq 0$ if $f$ is continuous on $\mathcal{X} \subseteq \mathbb{R}^n$ and all the partial derivatives of order $1,\hdots,k$ exist and are continuous on $\mathcal{X}$, and $f$ is \emph{piecewise-}$\mathscr{C}^k$ with $k \geq 0$ if there exists a partition of $\mathcal{X}$ such that $f$ is $\mathscr{C}^k$ on each set in the partition.

\subsection{Interval Analysis}\label{sec:prelem-interval-analysis}
We denote the set of intervals on $\mathbb{R}$ by $\mathbb{IR} = \{ \mathcal{A} = [\underline{\mathcal{A}},\overline{\mathcal{A}}]  \: | \: \underline{\mathcal{A}},\overline{\mathcal{A}} \in \mathbb{R}, \underline{\mathcal{A}} \leq \overline{\mathcal{A}}\}$, the set of $n$-dimensional interval vectors by $\mathbb{IR}^n$, and the set of $n\times m$-dimensional interval matrices by $\mathbb{IR}^{n \times m}$. We carry forward the definitions~\cite{moore1966interval} of arithmetic operations, set inclusion, and intersections of intervals to interval vectors and matrices by applying them componentwise. We use the term interval to specify an interval vector or interval matrix when it is clear from the context.

Given $f : \mathcal{X} \mapsto \mathcal{Y}$ with $\mathcal{X} \subseteq \mathbb{R}^n$ and $\mathcal{Y} \subseteq \mathbb{R}^{m}$ (or $\mathcal{Y} \subseteq \mathbb{R}^{n \times m}$), we define an \emph{interval extension of $f$} as $\boldsymbol{f}$ with
\begin{align}
     \boldsymbol{f}(\mathcal{A}) \supseteq \mathscr{R}(f, \mathcal{A}) = \{ f(x) \: | \: x \in \mathcal{A} \} , \quad \forall \mathcal{A} \subseteq \mathcal{X}. \label{eq:range-def-over-approximation}
\end{align}
Thus, given an interval $ \mathcal{A}$, $\boldsymbol{f}(\mathcal{A})$ is an interval that over-approximates the range of values taken by $f$ over $\mathcal{A}$.
\begin{example}[\textsc{Interval extension of $2-$norm}] \label{ex:norm-2-extension}
    Consider $f=||\cdot||_2$. We compute its interval extension $\boldsymbol{f}$ via interval extensions of $\alpha=\sqrt{\cdot}$ and $\beta={(\cdot)}^2$. For any $ \mathcal{A}=[\underline{\mathcal{A}}, \overline{\mathcal{A}}]\in \mathbb{IR}$,
    \begin{align}
        \boldsymbol{\alpha}(\mathcal{A}) &= [\sqrt{\underline{\mathcal{A}}}, \sqrt{\overline{\mathcal{A}}}], \: \: \: \text{if } \underline{\mathcal{A}} \geq 0 , \label{eq:sqrt_ext}\\
        \boldsymbol{\beta}(\mathcal{A}) &= \begin{cases} [0 , \max\{\underline{\mathcal{A}}^2,\overline{\mathcal{A}}^2\}],& \text{if } 0 \in \mathcal{A} \\ 
        [\min \{\underline{\mathcal{A}}^2,\overline{\mathcal{A}}^2\}, \max\{\underline{\mathcal{A}}^2,\overline{\mathcal{A}}^2\}],& \text{otherwise}.\end{cases}\label{eq:sqr_ext}
    \end{align}
    Using \eqref{eq:sqrt_ext} and \eqref{eq:sqr_ext} with interval arithmetic, we have, for any $ \mathcal{S}=[\mathcal{S}_1,\hdots, \mathcal{S}_n] \in \mathbb{IR}^n$, $\boldsymbol{f}( \mathcal{S}) = \boldsymbol{\alpha}\left({\sum_{i=1}^n \boldsymbol{\beta}( \mathcal{S}_i)}\right)$.
\end{example}

\subsection{Over-Approximations of the Reachable Set} \label{sec:prelem-taylor-method}

Consider a nonlinear dynamical system,
\begin{equation} \label{general_dynamic}
    \dot{x} = h (x , u),
\end{equation}
where the state $x : \mathbb{R}_+ \mapsto \mathcal{X}$ is a continuous-time signal evolving in $\mathcal{X} \in \mathbb{IR}^n$, the control $u \in \mathbb{U}$ is a signal of time evolving in the control set $\mathcal{U}\in\mathbb{IR}^m$ with $\mathbb{U}=\{v : \mathbb{R}_+ \mapsto \mathcal{U}  \: | \: v \text{ is \emph{piecewise-}}\mathscr{C}^{D_u} \}$ for $D_u \geq 0$, and $h : \mathcal{X} \times \mathcal{U} \mapsto \mathcal{Y}$ is $\mathscr{C}^{D_h}$ for $D_h\geq 1$ and $\mathcal{Y} \in \mathbb{IR}^n$. Given an initial state $x_i = x(t_i)$ at time $t_i$ and a control signal $u \in \mathbb{U}$, a trajectory of~\eqref{general_dynamic} is a function of time $x(\cdot;x_i, u) : [t_i,\infty[ \mapsto \mathcal{X}$ that satisfies~\eqref{general_dynamic}.
\begin{definition} [\textsc{Reachable set}]
    Given a set $\mathcal{I}_i \subseteq \mathcal{X}$ of states  at time $t_i$ and a set $\mathbb{V} \subseteq \mathbb{U}$ of control signals, the reachable set of the dynamics~\eqref{general_dynamic} at time $t\geq t_i$ is given by
    \begin{equation*}
        \mathcal{R}(t, \mathcal{I}_i,\mathbb{V}) =\{ z \in  \mathcal{X} \: \vert \:  \exists x_i \in \mathcal{I}_i, \: \exists v \in \mathbb{V}, \: z=x(t;x_i,v)\}.
    \end{equation*}
\end{definition}

Given a set $\mathbb{V} \subseteq \mathbb{U}$ and a set $\mathcal{I}_0 \subseteq \mathcal{X}$ of states at time $t_0$, we compute over-approximations of $\mathcal{R}(t, \mathcal{I}_0,\mathbb{V})$ at time $t \geq t_0$ using interval Taylor-based methods~\cite{moore1966interval,nedialkov1999validated,goubault2019inner}. Specifically, we consider a time grid $t_0 <\cdots <t_N$ such that for all $v \in \mathbb{V}$, $v \text{ is } \mathscr{C}^{D_u}$ on each interval $[t_i, t_{i+1})$. We want to compute sets $ \mathcal{R}_i^+\in \mathbb{IR}^n$ (with $\mathcal{R}_0^+=\mathcal{I}_0$) such that for all $i \in \mathbb{N}_{[0,N-1]}$, $\mathcal{R}_{i+1} = \mathcal{R}(t_{i+1}, \mathcal{R}_{i}^+,\mathbb{V}) \subseteq \mathcal{R}_{i+1}^+$.
First, interval arithmetic enables to inductively define $\boldsymbol{h}^{[d]}$, the interval extensions of the Taylor coefficients $h^{[d]}$ given by
\begin{align}
    h^{[1]} = h, \: h^{[d+1]} = \frac{1}{d+1}\Big( \frac{\partial h^{[d]}}{\partial x} h + \sum_{l=0}^{d-1} \frac{\partial h^{[d]}}{\partial u ^{(l)}} u^{(l+1)} \Big). \label{eq:taylor-coeff}
\end{align}
Next, we start with $\mathcal{R}_0^+= \mathcal{I}_0$, and compute $\{\mathcal{R}_{i}^+\}_{i=1}^N$ by
\begin{align}
    \mathcal{R}_{i+1}^+= \mathcal{R}_{i}^+ & + \sum_{d=1}^{D-1} \big( t_{i+1}-t_{i}\big)^d \big( \boldsymbol{h}^{[d]}(\mathcal{R}_{i}^+, \boldsymbol{v})\big) (t_i) \nonumber \\
                     &+ \big(t_{i+1}-t_{i}\big)^{D} \big( \boldsymbol{h}^{[D]}(\mathcal{S}_{i},\boldsymbol{v})\big)([t_i,t_{i+1}]), \label{eq:taylor-expansion}
\end{align}
where $D\leq\min(D_u+1,D_h)$ is the order of the Taylor expansion, we denote $\boldsymbol{v}^{(0)}(\mathcal{A})$ by $\boldsymbol{v}(\mathcal{A})$ and the intervals $\boldsymbol{v}^{(d)}(\mathcal{A})$ for all $d\in \mathbb{N}_{[0,D_u]}$ are such that $\cup_{v \in \mathbb{V}} \mathscr{R}(v^{(d)},\mathcal{A}) \subseteq \boldsymbol{v}^{(d)}(\mathcal{A})$ with $\mathcal{A} \subseteq \mathbb{R}_+$. In other words, $\boldsymbol{v}^{(d)}(\mathcal{A})$ over-approximates the range of the $d^{\mathrm{th}}$  derivative of all $v \in \mathbb{V}$ on the interval $\mathcal{A}$. Here, the set $\mathcal{S}_{i} \subseteq \mathcal{X}$ is an \emph{a priori rough enclosure} of $\mathcal{R}(t, \mathcal{R}_i^+, \mathbb{V})$ for all $t \in [t_i, t_{i+1}]$ and is a solution of
\begin{align}
    \mathcal{R}_{i}^+ \ + \  [0, t_{i+1}-t_{i}]\ \boldsymbol{h}(\mathcal{S}_{i},\boldsymbol{v}([t_i,t_{i+1}])) \ \subseteq \ \mathcal{S}_{i}. \label{eq:rough-enclosure-approx}
\end{align}

\subsection{Problem Statement} \label{sec:problem-statement}

In this paper, we consider \emph{control-affine} nonlinear dynamics,
\begin{align}
    \dot{x} = f(x) + G(x) u, \label{eq:control-linearization}
\end{align}
where $f : \mathcal{X} \mapsto \mathbb{R}^n$ and $G : \mathcal{X} \mapsto \mathbb{R}^{n \times m}$ are \emph{unknown} vector-valued and matrix-valued functions, respectively, for $\mathcal{X} \in \mathbb{IR}^n$. Note that even though we consider control-affine dynamics, in the general case, we can construct a control-affine model of the system locally and apply the results of the paper.

\begin{assumption}[\textsc{Smoothness}]\label{assum:smooth}
    $f$ and $G$  are $\mathscr{C}^\mathrm{D}$ functions for some $D\geq 1$.
\end{assumption}
Assumption~\ref{assum:smooth} is common in the frameworks of reachability analysis and receding-horizon control. It implies that $f$ and $G$ are globally Lipschitz-continuous since the domain $\mathcal{X} \in \mathbb{IR}^n$ is bounded. We exploit such Lipschitz continuity by considering the following side information throughout the paper.

\begin{sideinfo}[\textsc{Bounds on Lipschitz constants}]\label{side:Lip}
Let $L_f \in  \mathbb{R}_+^n$ and $L_G \in \mathbb{R}_+^{n \times m}$ with $(L_f)_k = L_{f_k}$ and $(L_G)_{k,l} = L_{G_{k,l}}$ be known upper bounds on the Lipschitz constants of $(f)_k$ and $(G)_{k,l}$ for all $k \in \mathbb{N}_{[1,n]}$ and $l \in \mathbb{N}_{[1,m]}$.
\end{sideinfo}

We can also utilize, if available, any of the following side knowledge on the underlying dynamics.
\begin{sideinfo}[\textsc{Vector field bounds}]\label{side:bounds-state-vectorfield}
We are given  $\mathcal{R}^{f_\mathcal{A}} \in \mathbb{IR}^n$ and $\mathcal{R}^{G_{\mathcal{A}}} \in \mathbb{IR}^{n \times m}$ as known over-approximations of the range of $f$ and $G$, respectively, over a given set $\mathcal{A} \subseteq \mathcal{X}$.
\end{sideinfo}

\begin{sideinfo}[\textsc{Gradient bounds}]\label{side:gradient-bounds}
    We are given bounds on the gradient of some components of $f$ and $G$. Such side information may include the monotonicity of $f$ and $G$.
\end{sideinfo}

\begin{sideinfo}[\textsc{Decoupling among states}] \label{side:state-dependency}
    We are given the knowledge that some components of the vector field $\dot{x}$ do not depend on some components of the state $x$.
\end{sideinfo}

\begin{sideinfo}[\textsc{Partial dynamics knowledge}]\label{side:partial-knowledge}
    We are given terms of some components of the vector field as $$\dot{x} = f_{\mathrm{kn}}(x) + f_{\mathrm{ukn}}(x) + (G_{\mathrm{kn}}(x) + G_{\mathrm{ukn}}(x)) u,$$ where $f_{\mathrm{kn}}$ and $G_{\mathrm{kn}}$ are known functions while $f_{\mathrm{ukn}}$ and $G_{\mathrm{ukn}}$ are unknown functions. Such side information can be derived from the application of elementary laws of physics.
\end{sideinfo}

Let $N \in \mathbb{N}$, $N\geq 1$. Let $\mathscr{T}_N = \{(x_i,\dot{x}_i, u_i)\}_{i=1}^{N}$ denote a single finite-horizon trajectory containing $N$ samples of the state $x_i=x(t_i)$, the derivative $\dot{x}_i=\dot{x}(t_i)$ of the state, and the control signal $u_i = u(t_i)$ from a trajectory of~\eqref{eq:control-linearization}. Given the current state $x(t_{N+1})$ and the trajectory $\mathscr{T}_N$, we first seek to over-approximate the reachable set of the system. 
\begin{problem}[\textsc{Reachable set over-approximation}] \label{prob:reachability}
    Given a single finite-horizon trajectory $\mathscr{T}_N$ for $N\in \mathbb{N}$, the side information~\ref{side:Lip} and possibly any of the side information~\ref{side:bounds-state-vectorfield}--\ref{side:partial-knowledge}, a set $\mathbb{V} \subseteq \mathbb{U}$ of admissible control signals, a time step size $\Delta t > 0$, and a maximum number  $T > N$ of time steps, compute an over-approximation of the reachable set at time $t_i = t_N + (i-N) \Delta t$ for all $ i \in \mathbb{N}_{[N+1,T]}$.
\end{problem}

Next, we seek to control the system by finding, at each sampling time $t_i > t_N$, control values solutions to the \emph{one-step optimal control} problem~\cite{kocijan2004gaussian,vinod2020convexified,park1999robust}
\begin{align}
    \underset{u_i \in \mathcal{U}}{\mathrm{minimize}}& \quad c(x_i, u_i, x(t_{i+1}; x_i, u_i)),
    \label{eq:one-step-optimal-control}
\end{align}
where $x_i$ is the state of the system at $t_i$, $t_{i+1} = t_i + \Delta t$, $\Delta t$ is the constant time step size, and $c:\mathcal{X}\times\mathcal{U}\times\mathcal{X}\to \mathbb{R}$ is a known convex function, which encodes the preferences over a short interval of length $\Delta t$ in time.
\begin{problem}[\textsc{Approximate one-step optimal control}] \label{prob:control}
    Given a single finite-horizon trajectory $\mathscr{T}_{i-1}$ for some $i\in \mathbb{N},i > N$, the current state $x_i$ at time $t_i$, the side information~\ref{side:Lip}, and possibly any of the side information~\ref{side:bounds-state-vectorfield}--\ref{side:partial-knowledge}, compute, at sampling time $t_i$, approximate solutions to the one-step optimal control problem~\eqref{eq:one-step-optimal-control} and characterize the suboptimality of such approximations.
\end{problem}

We use the following example throughout the paper.
\begin{example}[\textsc{Unicycle System}] \label{ex:unicycle-system}
    Consider the unicycle system with dynamics given by
    \begin{equation} \label{eq:unicycle-dynamic}
        \dot{p}_x = v \cos (\theta), \ \dot{p}_y = v \sin (\theta),\ \dot{\theta} = \omega,
    \end{equation}
    where the components of the state $x = [p_x, p_y, \theta]$ represent, respectively, the position in the $x$ plane, the $y$ plane, and the heading of the unicycle. The components of the control $u = [v , \omega]$ represent the speed and the turning rate, respectively. We consider the constraint set $\mathcal{U} = [-3,3]\times [-\pi,\pi]$. In the control-affine form~\eqref{eq:control-linearization}, we have $f = 0$, $(G)_{1,1} = \cos (\theta)$, $(G)_{2,1} = \sin (\theta)$, $(G)_{3,2} = 1$, and $(G)_{k,l} = 0$ otherwise. We assume the dynamics~\eqref{eq:unicycle-dynamic} are unknown and are given the loose Lipschitz bounds $L_f = [0.01,0.01,0.01]$, $L_{G_{1,1}} = L_{G_{2,1}} = 1.1, L_{G_{3,2}} = 0.1$, and $L_{G_{k,l}} = 0$ otherwise. Furthermore, we consider the knowledge that the vector field does not depend on its positions. That is, $f(x) = f(\theta)$ and $G(x) = G(\theta)$.
\end{example}

\section{Over-Approximations of the Reachable Set of Unknown Dynamical Systems}
We develop \reachalgo{} to address Problem~\ref{prob:reachability}. It constructs a data-driven differential inclusion that contains the vector field of the unknown dynamics using the side information~\ref{side:Lip}. Then, it utilizes an interval Taylor-based method to over-approximate the reachable set of dynamics described by the constructed differential inclusion.

\subsection{Differential Inclusion based on Lipschitz Continuity}
First, to aid the construction of the differential inclusion, we over-approximate $f$ and $G$ at each data point of $\mathscr{T}_N$. 
\begin{lemma}[\textsc{Contraction via data}] \label{lem:contraction}
    Consider a data point $(x_i,\dot{x}_i, u_i)$, an interval $\mathcal{F}_i \in \mathbb{IR}^n$ such that $f(x_i) \in \mathcal{F}_i$, and an interval $\mathcal{G}_i \in \mathbb{IR}^{n \times m}$ such that $G(x_i) \in \mathcal{G}_i$. Let $C_{\mathcal{F}_i}$ and $C_{\mathcal{G}_i}$ be defined sequentially for $l=\mathbb{N}_{[1,m]}$ and $k\in \mathbb{N}_{[1,n]}$ by
    \begin{equation} \label{eq:contraction-fG}
        \begin{aligned}
            (C_{\mathcal{F}_i})_k &= (\mathcal{F}_i)_k \: \cap \: (\dot{x}_i - \mathcal{G}_i u_i)_k, \\
            (\mathcal{S}_0)_k &=  (\dot{x}_i - C_{\mathcal{F}_i})_k \cap (\mathcal{G}_i u_i)_k, \\
            (C_{\mathcal{G}_i})_{k,l} &= \begin{cases} ( ((\mathcal{S}_{l-1})_k - \sum_{p>l} (\mathcal{G}_i)_{k,p} (u_i)_p) \cap  ((\mathcal{G}_i)_{k,l} u_i^l) ) \frac{1}{u^l_i},\\
                                                \quad \text{if } u^l_i = (u_i)_l \neq 0 \\
                                                (\mathcal{G}_i)_{k,l}, \quad \text{otherwise},
                                    \end{cases}\\
            (\mathcal{S}_l)_k &= \big((\mathcal{S}_{l-1})_k - (C_{\mathcal{G}_i})_{k,l} (u_i)_l) \big) \cap \big( \sum_{p > l} (\mathcal{G}_i)_{k,p} (u_i)_p \big).
        \end{aligned}
    \end{equation}
    Then, $C_{\mathcal{F}_i}$ and $C_{\mathcal{G}_i}$ are the smallest intervals enclosing $f(x_i)$ and $G(x_i)$, respectively, given only the data point, $\mathcal{F}_i$, and $\mathcal{G}_i$.
\end{lemma}

We provide a proof for Lemma~\ref{lem:contraction} in \myextended{} where we use the constraint $\dot{x}_i = f(x_i) + G(x_i) u_i$ to remove from $\mathcal{F}_i$ and $\mathcal{G}_i$ values of $f(x_i)$ and $G(x_i)$ not satisfying the constraint.

Next, we use the Lipschitz continuity to provide explicit functions that over-approximate $f$ and $G$, given uncertain knowledge of $f$ and $G$ at the data points.
\begin{lemma}[\textsc{Over-approximation of $f$ and $G$}]\label{lem:overapprox-f-G}
    Given a set $\mathscr{E}_N = \{ (x_i,C_{\mathcal{F}_i}, C_{\mathcal{G}_i})\: | \: f(x_i) \in C_{\mathcal{F}_i}, G(x_i) \in C_{\mathcal{G}_i}\}_{i=0}^{N}$ and the bounds $L_f$ and $L_G$ from side information~\ref{side:Lip}, the interval-valued functions $\boldsymbol{f} : \mathcal{X} \to \mathbb{IR}^{n}$ and $\boldsymbol{G} : \mathcal{X} \to \mathbb{IR}^{n \times m}$, given for all $k\in \mathbb{N}_{[1,n]}$ and $l \in \mathbb{N}_{[1,m]}$ by
    \begin{equation}\label{eq:overapprox-f-G}
    \begin{aligned}
        (\boldsymbol{f}(x))_k &=  \bigcap_{(x_i, C_{\mathcal{F}_i}, \cdot) \in \mathscr{E}_N} (C_{\mathcal{F}_i})_k + L_{f_k} \| x - x_i\|_2 [-1,1], \\
        (\boldsymbol{G}(x))_{k,l} &= \bigcap_{(x_i, \cdot, C_{\mathcal{G}_i}) \in \mathscr{E}_N} (C_{\mathcal{G}_i})_{k,l} + L_{G_{k,l}} \| x - x_i\|_2 [-1,1],
    \end{aligned}
    \end{equation}
    are such that $f(x) \in \boldsymbol{f}(x)$ and $G(x) \in \boldsymbol{G}(x)$ for all $x \in \mathcal{X}$.
\end{lemma}
We provide a proof for Lemma~\ref{lem:overapprox-f-G} in \myextended{}. Note that interval extensions of $\boldsymbol{f}$ and $\boldsymbol{G}$ can be obtained by replacing occurences of $\|\cdot\|_2$ by its interval extension given in Example~\ref{ex:norm-2-extension}. In the rest of the paper, when the input of $\boldsymbol{f}$ and $\boldsymbol{G}$ are assumed to be intervals, we use such interval extensions.

Finally, Theorem~\ref{thm:diff-inclusion} constructs a differential inclusion for the system by combining Algorithm~\ref{algo:overapprox-datapoints} and Lemma~\ref{lem:overapprox-f-G}.

\algdef{SE}[DOWHILE]{Do}{doWhile}{\algorithmicdo}[1]{\algorithmicwhile\ #1}%
\begin{algorithm}[!t]
    \caption{Over-approximation of the values of $f$ and $G$ at each data point of a finite-horizon trajectory.}\label{algo:overapprox-datapoints}
    \begin{algorithmic}[1]    
        \Require{Single trajectory $\mathscr{T}_N$, sufficiently large $M >0$, upper bounds on the Lipschitz constants (side information~\ref{side:Lip}).\newline
        \emph{Optional}: The sets $\mathcal{A}, \mathcal{R}^{f_\mathcal{A}}$ and $ \mathcal{R}^{G_\mathcal{A}}$ (side information~\ref{side:bounds-state-vectorfield}).}
    \Ensure{${\mathscr{E}_N = \{ (x_i,C_{\mathcal{F}_i}, C_{\mathcal{G}_i}) |  f(x_i) \in C_{\mathcal{F}_i}, G(x_i) \in C_{\mathcal{G}_i}\}_{i=0}^{N}}$} 
        \If{no side information~\ref{side:bounds-state-vectorfield}}
            \State $\mathcal{A} \gets \mathcal{X}$, $\mathcal{R}^{f_\mathcal{A}} \gets [-M,M]^n$, $\mathcal{R}^{G_\mathcal{A}} \gets [-M,M]^{n \times m}$ 
        \EndIf
        \State Define $x_0 \in \mathcal{A}$, $C_{\mathcal{F}_0} \gets \mathcal{R}^{f_\mathcal{A}}$, and $C_{\mathcal{G}_0} \gets \mathcal{R}^{G_\mathcal{A}}$
        \For{$i \in \mathbb{N}_{[1,N]} \wedge (x_i, \dot{x}_i, u_i) \in \mathscr{T}_{N}$} \label{alg:begin-init-e}
                \State Compute $\mathcal{F}_i = \boldsymbol{f}(x_i), \mathcal{G}_i = \boldsymbol{G}(x_i)$ via~\eqref{eq:overapprox-f-G} and $\mathscr{E}_{i-1}$ \label{alg:update-ei}
                \State Compute $C_{\mathcal{F}_i}, C_{\mathcal{G}_i}$ via~\eqref{eq:contraction-fG}, $\mathcal{F}_i, \mathcal{G}_i$, and $(x_i,\dot{x}_i,u_i)$ \label{alg:contraction-f-G}
        \EndFor \label{alg:end-init-e}
        \While {$\mathscr{E}_N$ is not invariant} \label{alg:while-begin}
            \State Execute lines~\ref{alg:begin-init-e}--\ref{alg:end-init-e} with $\mathscr{E}_N$ instead of $\mathscr{E}_{i-1}$ in line~\ref{alg:update-ei} 
        \EndWhile \label{alg:while-end}
    \State \Return $\mathscr{E}_N$
  \end{algorithmic}
\end{algorithm}

\begin{theorem}[\textsc{Differential inclusion}]\label{thm:diff-inclusion}
    Given a trajectory $\mathscr{T}_N$, the bounds $L_f$ and $L_G$ from side information~\ref{side:Lip}, and possibly bounds on the vector field from side information~\ref{side:bounds-state-vectorfield}, the dynamics~\eqref{eq:control-linearization} are contained in the differential inclusion
    \begin{align}
        \dot{x} \in \boldsymbol{f}(x) + \boldsymbol{G}(x) u, \label{eq:diff-inclusion}
    \end{align}
    where the functions $\boldsymbol{f} : \mathcal{X} \to \mathbb{IR}^{n}$ and $\boldsymbol{G} : \mathcal{X} \to \mathbb{IR}^{n \times m}$ are obtained by~\eqref{eq:overapprox-f-G} with $\mathscr{E}_N$ being the output of Algorithm~\ref{algo:overapprox-datapoints}.
\end{theorem}

A proof of Theorem~\ref{thm:diff-inclusion} is provided in \myextended{} where we show that the output $\mathscr{E}_N$ of Algorithm~\ref{algo:overapprox-datapoints} is such that for all $(x_i, C_{\mathcal{F}_i}, C_{\mathcal{G}_i}) \in \mathscr{E}_N$, $f(x_i) \in C_{\mathcal{F}_i}$ and $G(x_i) \in C_{\mathcal{G}_i}$. Therefore, Lemma~\ref{lem:overapprox-f-G} and interval arithmetic enable to conclude.

\begin{figure}[!hbt]
    \vspace*{-1.5mm}
    \centering
    \hspace*{-0.4cm}
    \input{derpx_unicycle_evolution}
    \vspace*{-3mm}
    \caption{Time evolution of $\dot{p}_x$ and its over-approximation $(\boldsymbol{f}(x) + \boldsymbol{G}(x) u)_1$ for the unicycle system of Example~\ref{ex:unicycle-system}. We generate the trajectory corresponding to $\dot{x}(t)$ using a randomly generated piecewise-constant input $u(t) \in \mathcal{U}$ for $t \leq t_{15} = 1.5s$, and $u(t) = [1 , \cos(6(t-t_{15}))]$ for $t \in [t_{15}, 4]$.}
    \vspace*{-3mm}
    \label{fig:diff-inclusion-unicycle}
\end{figure}
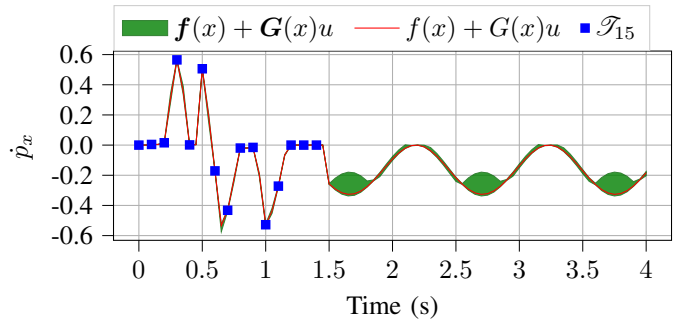

Figure~\ref{fig:diff-inclusion-unicycle} shows that the differential inclusion holds on the unicycle system using a randomly generated trajectory $\mathscr{T}_{15}$.

\begin{remark}
    The quality of the differential inclusion of Theorem~\ref{thm:diff-inclusion} depends on how much information on $f$ and $G$ can be obtained from the given trajectory $\mathscr{T}_N$. Thus, if the trajectory is not \emph{diverse}, it is likely impossible to obtain tight differential inclusions. For example, consider one of the corner cases where $u_i =0$ for every data point in $\mathscr{T}_N$. In this case, it is impossible to retrieve any information on $G$. When the goal is to control the unknown system, persistent excitations can help to diversify the trajectory and obtain tight differential inclusions, which ultimately improve future control decisions.
\end{remark}

\subsection{ Interval Taylor-Based Method for Differential Inclusions}

We compute an over-approximation of the reachable set of the dynamics described by the differential inclusion~\eqref{eq:diff-inclusion}. Such over-approximation is naturally an over-approximation of the reachable set of the unknown dynamical system.
\begin{theorem}[\textsc{Reachable set over-approximation}] \label{thm:over-approximation-state}
    Given a trajectory $\mathscr{T}_N$, a set $\mathbb{V} \subseteq \mathbb{U}$ of \emph{piecewise-}$\mathscr{C}^{D_u}$ control signals for $D_u \geq  1$, the bounds $L_f$ and $L_G$ from side information~\ref{side:Lip}, the set $\mathcal{R}^+_i \in \mathbb{IR}^n$ of states at time $t_i$, and a time step size $\Delta t$, an over-approximation $\mathcal{R}^+_{i+1}$ of the reachable set at $t_i+ \Delta t$ of dynamics described by the differential inclusion $\dot{x}\in \boldsymbol{f}(x) + \boldsymbol{G}(x) u$ with $u\in\mathbb{V}$ is given by
    \begin{align} 
        \mathcal{R}^+_{i+1} = \begin{aligned}[t] 
                                    \mathcal{R}^+_i &+  \Big( \boldsymbol{f}(\mathcal{R}^+_i) + \boldsymbol{G} (\mathcal{R}^+_i) \boldsymbol{v}(t_i) \Big) \Delta t  \\
                                    &+  \Big( \mathcal{J}_f + \mathcal{J}_G \mathcal{V}_i \Big) \Big( \boldsymbol{f}(\mathcal{S}_i) + \boldsymbol{G}(\mathcal{S}_i) \mathcal{V}_i \Big) \frac{\Delta t^2}{2} \\
                                    &+  \boldsymbol{G}(\mathcal{S}_i) \mathcal{V}^{(1)}_i \frac{\Delta t^2}{2},
                                \end{aligned} \label{eq:over-approx-next-state}
    \end{align}
    where $\mathcal{V}_i = \boldsymbol{v}([t_i, t_i+\Delta t])$, $\mathcal{V}^{(1)}_i = \boldsymbol{v}^{(1)}([t_i, t_i+\Delta t])$, and the matrices $\mathcal{J}_f \in \mathbb{IR}^{n \times n}$ and $\mathcal{J}_G \in \mathbb{IR}^{n \times m \times n}$, interval extensions of the Jacobian of $f$ and $G$, are such that $(\mathcal{J}_{f})_{k,p} = L_{f_k} [-1,1]$ and $(\mathcal{J}_G)_{k,l,p} = L_{G_{k,l}} [-1,1]$ for all $k,p \in \mathbb{N}_{[1,n]}$ and $l \in \mathbb{N}_{[1,m]}$. Besides, the set $\mathcal{S}_i$ can be obtained by solving
    \begin{equation} \label{eq:fix-point-rough}
        \mathcal{R}^+_i \ +  \ [0, \Delta t] \ \Big( \boldsymbol{f}(\mathcal{S}_i) + \boldsymbol{G}(\mathcal{S}_i) \mathcal{V}_i \Big) \ \subseteq \ \mathcal{S}_i.
    \end{equation}
    For $k,p\in \mathbb{N}_{[1,n]}$, we define the $(k,p)$ component of $\mathcal{J}_G \mathcal{V}_i$ as
    \begin{align}
        {(\mathcal{J}_G \mathcal{V}_i)}_{k,p} &= \sum_{l=1}^m
                {(\mathcal{J}_G)}_{k,l,p}
        {(\mathcal{V}_i)}_l. \label{eq:over-approx-jac-Gu}
    \end{align}
\end{theorem}

We provide a proof for Theorem~\ref{thm:over-approximation-state} in \myextended{} where we exploit a Taylor expansion~\eqref{eq:taylor-expansion} of order $D=2$ to derive such a result.

\begin{corollary}\label{corr:non-c1-control}
    Under the notation of Theorem~\ref{thm:over-approximation-state}, assume that $D_u = 0$. Then, an over-approximation $\mathcal{R}_{i+1}^+$ of the reachable set of the dynamics described by the differential inclusion $\dot{x}\in \boldsymbol{f}(x) + \boldsymbol{G}(x) u$ of Theorem~\ref{thm:diff-inclusion} with $u\in\mathbb{V}$ is given by
    \begin{align*}
        \mathcal{R}_{i+1}^+ = \mathcal{R}_i^+ + \big( \boldsymbol{f}(\mathcal{S}_i) + \boldsymbol{G}(\mathcal{S}_i) \mathcal{V}_i \big) \Delta t.
    \end{align*}
\end{corollary} 

Theorem~\ref{thm:over-approximation-state}, as it is, incorporates only the Lipschitz bounds from side information~\ref{side:Lip}. We now show how to incorporate the other side information  to obtain tighter over-approximations.\vspace*{1mm}

\noindent {\bf{Side information~\ref{side:bounds-state-vectorfield}}} (\textsc{Vector field bounds}). Given a set $\mathcal{S} \subseteq \mathcal{A}$, tighter interval extensions of $\boldsymbol{f}$ and $\boldsymbol{G}$ over $\mathcal{S}$ can be obtained by the update $\boldsymbol{f}(\mathcal{S}) \gets \boldsymbol{f}(\mathcal{S}) \cap \mathcal{R}^{f_\mathcal{A}}$ and $\boldsymbol{G}(\mathcal{S}) \gets \boldsymbol{G}(\mathcal{S}) \cap \mathcal{R}^{G_\mathcal{A}}$.\vspace*{2mm}

\noindent {\bf{Side information~\ref{side:gradient-bounds}}} (\textsc{Gradient bounds}). These bounds can be used to provide tighter interval extensions $\mathcal{J}_f$ and $\mathcal{J}_G$ of the Jacobians of $f$ and $G$. For example, if the function $(f)_k$ is known to be non-decreasing with respect to the variable $x_p$ on a set $\mathcal{A} \subseteq \mathcal{X}$, then we obtain a tighter $\mathcal{R}_{i+1}^+$ by the update $(\mathcal{J}_f)_{k,p} \gets (\mathcal{J}_f)_{k,p} \cap \mathbb{R}_+$ if $\mathcal{S}_i \subseteq \mathcal{A}$.\vspace*{2mm}

\noindent {\bf{Side information~\ref{side:state-dependency}}} (\textsc{Decoupling among states}). This knowledge constrains the interval extensions for the Jacobian matrices. For example, if the state ${(x(t))}_{p}$ does not directly affect ${(\dot{x}(t))}_{k}$ for some $p,k \in \mathbb{N}_{[1,n]}$ under any control signal in $ \mathcal{U}$, we can obtain a tighter over-approximation of the reachable set by setting to zero the intervals ${(\mathcal{J}_f)}_{k,p}$ and ${(\mathcal{J}_G)}_{k, l, p}$ for all $l \in \mathbb{N}_{[1,m]}$. For the unicycle system of Example~\ref{ex:unicycle-system}, since $f$ and $G$ depend only on the heading $\theta$, the Jacobian terms $(\mathcal{J}_f)_{1,1}$, $(\mathcal{J}_f)_{1,2}$, $(\mathcal{J}_f)_{2,1}$, $(\mathcal{J}_f)_{2,2}$, $(\mathcal{J}_f)_{3,1}$, $(\mathcal{J}_f)_{3,2}$, ${(\mathcal{J}_G)}_{1,1,1}$, ${(\mathcal{J}_G)}_{1,1,2}$, ${(\mathcal{J}_G)}_{2,1,1}$, ${(\mathcal{J}_G)}_{2,1,2}$, ${(\mathcal{J}_G)}_{3,2,1}$, and ${(\mathcal{J}_G)}_{3,2,2}$ must all be set to \emph{zero}.\vspace*{2mm}

\noindent {\bf{Side information~\ref{side:partial-knowledge}}} (\textsc{Partial dynamics knowledge}). The new functions $\boldsymbol{f}$ and $\boldsymbol{G}$ are given by $\boldsymbol{f} = \boldsymbol{f}_{\mathrm{kn}} + \boldsymbol{f}_{ukn}$ and $\boldsymbol{G} = \boldsymbol{G}_{\mathrm{kn}} + \boldsymbol{G}_{ukn}$, where the functions $\boldsymbol{f}_{\mathrm{kn}}$ and $\boldsymbol{f}_{ukn}$ are interval extensions of known functions $f_{\mathrm{kn}}$ and $G_{\mathrm{kn}}$, respectively, and the functions $\boldsymbol{f}_{ukn}$ and $\boldsymbol{G}_{ukn}$ are obtained by Theorem~\ref{thm:diff-inclusion} using $L_{f_{\mathrm{ukn}}}$, $L_{G_{\mathrm{ukn}}}$, and the new trajectory $$\mathscr{T}^{'}_{N} = \{(x_i,\dot{x}_i - (f_{\mathrm{kn}}(x_i) +G_{\mathrm{kn}}(x_i) u_i) , u_i) \: \vert \: (x_i, \dot{x}_i, u_i) \in \mathscr{T}_N\}.$$
Furthermore, the new Jacobian terms in the computation of $\mathcal{R}_{i+1}^+$ are given by $\mathcal{J}_f = \boldsymbol{\frac{\partial f_{\mathrm{kn}}}{\partial x}}(\mathcal{S}_i) + \mathcal{J}_{f_{\textrm{ukn}}}$ and $(\mathcal{J}_{G})_{k,l,p} = \boldsymbol{\frac{\partial (G_{\mathrm{kn}})_{k,l}}{\partial x_p}}(\mathcal{S}_i) + (\mathcal{J}_{G_{\textrm{ukn}}})_{k,l,p}$, respectively.\vspace*{1mm}

\algdef{SE}[DOWHILE]{Do}{doWhile}{\algorithmicdo}[1]{\algorithmicwhile\ #1}%
\begin{algorithm}[t]
    \caption{\reachalgo{}: Over-approximation of the reachable set of unknown smooth systems.}\label{algo:DaTaReach}
    \begin{algorithmic}[1]    
        \Require{Single trajectory $\mathscr{T}_N$, upper bounds on the Lipschitz constants from side information~\ref{side:Lip}, set $\mathbb{V}$ of control signals, time step size $\Delta t$, maximum number of time steps $T > N$.\newline
        \emph{Optional}: Any of the side information~\ref{side:bounds-state-vectorfield}--\ref{side:partial-knowledge}.}
    \Ensure{Over-approximations ${\{\mathcal{R}^+_i\}}_{i = N+1}^T$ of the reachable sets at times $t_N + (i-N)\Delta t$ with $i \in \mathbb{N}_{[N+1,T]}$.} 
        \State Define $R_{N+1}^+\gets\{x(t_{N+1})\}$
        \State Compute $\boldsymbol{f}$, and $\boldsymbol{G}$ from Theorem~\ref{thm:diff-inclusion}
        \For{$i\in\{N+1,\ldots,T-1\}$}
                \State Compute $\boldsymbol{v}(t_i)$, $\mathcal{V}_i$, and $\mathcal{V}^{(1)}_i$ in Theorem~\ref{thm:over-approximation-state} from $\mathbb{V}$
                \State Compute $\mathcal{S}_i$ via~\eqref{eq:fix-point-rough}, $\mathcal{V}_i$, $\boldsymbol{f}$, and $\boldsymbol{G}$
                \State Compute $\mathcal{R}^+_{i+1}$ via~\eqref{eq:over-approx-next-state},  $\mathcal{R}^+_i$,  $\mathcal{S}_i$, $\boldsymbol{f}$, $\boldsymbol{G}$, $\mathcal{J}_f$, and $\mathcal{J}_G$
        \EndFor
    \State\Return ${\{\mathcal{R}^+_i\}}_{i=N+1}^T$
  \end{algorithmic}
\end{algorithm}

\begin{figure}[!hbt]
    \centering
    \hspace*{-0.5cm}
    \input{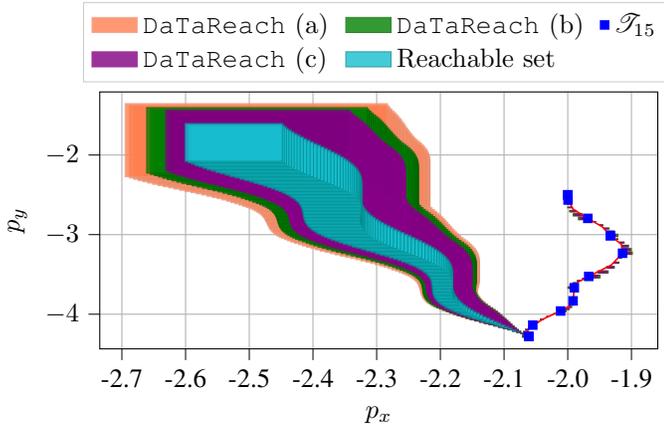}
    \vspace*{-4mm}
    \caption{Over-approximation of the reachable set of the unicycle system in the $x$-$y$ plane. The trajectory $\mathscr{T}_{15}$ is obtained by a randomly generated piecewise-constant control signal $u(t) \in \mathcal{U}$ for $t \leq t_{15} = 1.5s$. The parameters for \reachalgo{} were given by $\Delta t = 0.02$, $T = 200$, and $\mathbb{V} = \{t \mapsto [1+ a_1, \cos(6(t-t_{15})) + a_2] \: \vert \: a_1 \in [-0.1,0.1], a_2 \in [-0.01,0.01]\}$. The case $\mathrm{(a)}$ ignores the side information $f(x) = f(\theta)$, case $\mathrm{(b)}$ considers exactly the setting of Example~\ref{ex:unicycle-system}, and case $\mathrm{(c)}$ uses the same knowledge as in case $\mathrm{(b)}$ while assuming the extra knowledge that $f=0$ and $(G)_{3,2} = 1$.}
    \vspace*{-3mm}
    \label{fig:overapprox-unicycle}
\end{figure}

We demonstrate the value of additional side information on the unicycle system of Example~\ref{ex:unicycle-system}. Specifically, we use \reachalgo{} to compute over-approximations of the reachable sets of the unicycle under different side information. Figure~\ref{fig:overapprox-unicycle} shows that, as expected, the over-approximation becomes tighter under additional side information.

\section{Control Synthesis for Unknown Dynamical Systems}

We develop \controlalgo{} to address Problem~\ref{prob:control}. It computes approximate solutions to the one-step optimal control problem~\eqref{eq:one-step-optimal-control} using the over-approximation of the reachable set. Specifically, it replaces the \emph{unknown} next state $x_{i+1}$ in problem~\eqref{eq:one-step-optimal-control} with the corresponding over-approximation $\mathcal{R}^+_{i+1}$, which may be computed using Theorem~\ref{thm:over-approximation-state}. Unfortunately, $\mathcal{R}^+_{i+1}$ is a nonconvex function of the control signal $u$, the decision variable. Therefore, it constructs an over-approximation of $\mathcal{R}^+_{i+1}$, in Theorem~\ref{thm:linear-expr-O}, that is convex in the constant control signal $u$ applied during the time interval $[t_i,t_{i+1}]$. The new over-approximation provides tractable convex relaxations of the one-step optimal control problem \eqref{eq:one-step-optimal-control}.
\begin{theorem} \label{thm:linear-expr-O}
     Under the notation of Theorem~\ref{thm:over-approximation-state} and given a constant control signal $u : t \mapsto u_i$ applied between $t_i$ and $t_{i+1}=t_i + \Delta t$ with $u_i \in \mathcal{U}$, the reachable set $\mathcal{R}_{i+1}$ satisfies
    \begin{equation}
        \mathcal{R}_{i+1} \subseteq \big(\mathcal{B}_i + \mathcal{A}_i^+ u_i\big) \cap \big(\mathcal{B}_i + \mathcal{A}_i^- u_i\big), \label{eq:linear-over-approx}
    \end{equation}
    where the intervals $\mathcal{A}^-_i$, $\mathcal{A}^+_i$, and $\mathcal{B}_i$ are given by
    \begin{equation} \label{eq:linear-over}
    \begin{aligned}
        \mathcal{A}^-_i &= 
                                \boldsymbol{G}(\mathcal{R}^+_i) \Delta t + \big( \mathcal{J}_f \boldsymbol{G}(\mathcal{S}_i) + \mathcal{J}_G^{\mathrm{T}}( \boldsymbol{f}(\mathcal{S}_i) + \boldsymbol{G}(\mathcal{S}_i)\mathcal{U}) \big) \frac{\Delta t ^2}{2}, \\
        \mathcal{A}^+_i &= 
                                \boldsymbol{G}(\mathcal{R}^+_i) \Delta t + \big( (\mathcal{J}_f + \mathcal{J}_G \mathcal{U}) \boldsymbol{G}(\mathcal{S}_i) + \mathcal{J}_G^{\mathrm{T}} \boldsymbol{f}(\mathcal{S}_i) \big) \frac{\Delta t ^2}{2}, \\
        \mathcal{B}_i &= \mathcal{R}^+_i + \boldsymbol{f} (\mathcal{R}^+_i) \Delta t +   \mathcal{J}_f \boldsymbol{f}(\mathcal{S}_i) \frac{\Delta t^2}{2},
    \end{aligned}
    \end{equation}
    where ${(\mathcal{J}_G^\textrm{T})}_{k,p,l}$ = ${(\mathcal{J}_G)}_{k,l,p}$ for $k,p\in \mathbb{N}_{[1,n]}$ and $l \in \mathbb{N}_{[1,m]}$, and the a priori rough enclosure $\mathcal{S}_i$ is a solution of 
    \begin{equation}\label{eq:apriori-enclosure-control}
        \mathcal{R}^+_i \ + \ [0, \Delta t] \ \Big( \boldsymbol{f}(\mathcal{S}_{i}) + \boldsymbol{G}(\mathcal{S}_{i}) \mathcal{U} \Big) \ \subseteq \ \mathcal{S}_{i}.
    \end{equation}
\end{theorem}

A proof of Theorem~\ref{thm:linear-expr-O} is provided in \myextended{} where we linearize, using the constrained set $\mathcal{U}$, the quadratic term in $u$ of the closed-form expression~\eqref{eq:over-approx-next-state}. 

We use the control-affine linearization of Theorem~\ref{thm:linear-expr-O} to propose two convex optimization relaxations of the one-step optimal control problem~\eqref{eq:one-step-optimal-control}. The first relaxation, called \emph{optimistic control problem}, is given by
\begin{align}\label{eq:optimistic-control-problem}
    \underset{u_i \in \mathcal{U}}{\mathrm{minimize}} &\quad \underset{ \begin{smallmatrix} x_{i+1} \in \mathcal{X}, \\ x_{i+1} \in (\mathcal{B}_i + \mathcal{A}_i^+ u_i) \cap (\mathcal{B}_i + \mathcal{A}_i^- u_i) \end{smallmatrix}}{\inf} c(x_i, u_i, x_{i+1}), 
\end{align}
where the goal is to minimize the best possible cost value over all possible state $x_{i+1}$ in the over-approximation~\eqref{eq:linear-over-approx}. The second relaxation, called \emph{idealistic control problem}, is an idealistic approximation of~\eqref{eq:one-step-optimal-control} given by
\begin{align}\label{eq:idealistic-control-problem}
    \underset{u_i \in \mathcal{U}}{\mathrm{minimize}} &\quad  c(x_i, u_i, b^{\mathrm{ide}}_i + A^{\mathrm{ide}}_i u_i), 
\end{align}
where the goal is to minimize the cost associated to a specific trajectory $x_{i+1} = b^{\mathrm{ide}}_i + A^{\mathrm{ide}}_i u_i$ in the over-approximation~\eqref{eq:linear-over-approx}, idealistically considered as the unknown next state evolution.

\algdef{SE}[DOWHILE]{Do}{doWhile}{\algorithmicdo}[1]{\algorithmicwhile\ #1}%
\begin{algorithm}[!t]
    \caption{\controlalgo{}: Approximate at sampling time $t_i > t_N$ a solution to the one-step optimal control problem.}\label{algo:DaTaControl}
    \begin{algorithmic}[1]    
    \Require{Single trajectory $\mathscr{T}_{i-1}$, time step size $\Delta t$, upper bounds on the Lipschitz constants from side information~\ref{side:Lip}, one-step cost $c$, current state $x_i = x(t_{i})$, constraint set $\mathcal{U}$.\newline
        \emph{Optional}: Any of the side information~\ref{side:bounds-state-vectorfield}--\ref{side:partial-knowledge}.}
    \Ensure{Constant control $\hat{u}_i \in \mathcal{U}$ to apply between $t_i$ and $t_i + \Delta t$ that approximates a solution of~\eqref{eq:one-step-optimal-control}.} 
        \State Define $\mathcal{R}^+_i \leftarrow \{x_i\}$
        \State Compute $\boldsymbol{f}$, and $\boldsymbol{G}$ from Theorem~\ref{thm:diff-inclusion}
        \State Compute $\mathcal{S}_i$ via~\eqref{eq:apriori-enclosure-control}, $\mathcal{R}^+_i$, $\mathcal{U}$, $\boldsymbol{f}$, and $\boldsymbol{G}$
        \State Compute $\mathcal{B}_{i}$, $\mathcal{A}^+_{i}$, and $\mathcal{A}^-_{i}$ via~\eqref{eq:linear-over}, $\mathcal{R}^+_i$, $\mathcal{S}_i$, $\mathcal{U}$, $\boldsymbol{f}$, $\boldsymbol{G}$, $\mathcal{J}_f$, and $\mathcal{J}_G$
        \State Compute $\hat{u}_i$ as the solution of either \eqref{eq:optimistic-control-problem} or \eqref{eq:idealistic-control-problem}
        \State\Return $\hat{u}_i$
  \end{algorithmic}
\end{algorithm}

Finally, we provide, in Theorem~\ref{thm:suboptimality-gap}, bounds on the suboptimality of the relaxations when the cost $c$ is quadratic. 
\begin{figure*}[t]
    \centering
    \input{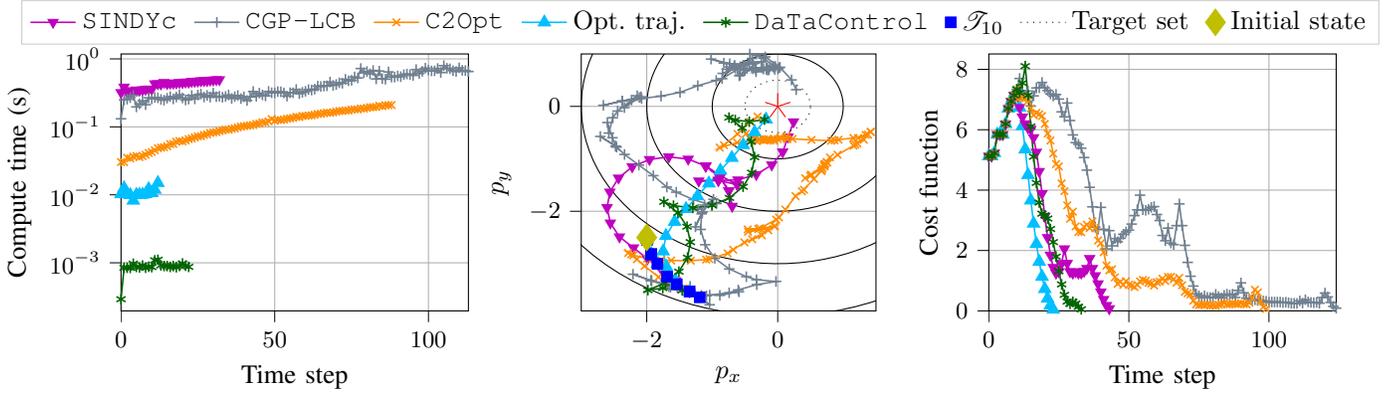}
    \vspace*{-5mm}
    \caption{\controlalgo{} successfully drove the unicycle to the origin. $\mathrm{Opt.\ traj.}$ corresponds to the one-step optimal control with the known dynamics.}
    \vspace*{-3.5mm}
    \label{fig:unicycle_traj_cost}
\end{figure*}

\begin{assumption}[\textsc{Quadratic one-step cost}]\label{ass:quadratic-cost}
    We consider that the one-step cost function $c$ is a convex quadratic function,
    \begin{align}
        c(x,u,y) = \begin{bmatrix} y \\ u\end{bmatrix}^{\mathrm{T}} \begin{bmatrix} Q & S \\ S^{\mathrm{T}}  & R\end{bmatrix} \begin{bmatrix} y \\ u\end{bmatrix} + \begin{bmatrix} q \\ r\end{bmatrix}^{\mathrm{T}} \begin{bmatrix} y \\ u\end{bmatrix},
    \end{align}
    where $q \in \mathbb{R}^n$, $r \in \mathbb{R}^m$, $Q = Q^{\mathrm{T}} \in \mathbb{R}^{n \times n}$, $R = R^{\mathrm{T}} \in \mathbb{R}^{m \times m}$, and $S \in \mathbb{R}^{n \times m}$.
\end{assumption}

\begin{theorem}[\textsc{Suboptimality bound}]\label{thm:suboptimality-gap}
    Let $c_i^{\star}$, $c_i^{\mathrm{opt}}$, and $c_i^{\mathrm{ide}}$ be the optimal cost of the one-step optimal control problem~\eqref{eq:one-step-optimal-control}, the optimistic control problem~\eqref{eq:optimistic-control-problem}, and the idealistic control problem~\eqref{eq:idealistic-control-problem}, respectively, at the sampling time $t_i$. We have
    \begin{align}
        |c_i^{\star} - c_i| &\leq \begin{aligned}[t] \max \big( &\| w(\mathcal{B}_i) + w(\mathcal{A}^+_i)|\mathcal{U}| \|_2 K (\mathcal{A}^+_i) , \\ 
        &\| w(\mathcal{B}_i) + w(\mathcal{A}^-_i)|\mathcal{U}| \|_2  K (\mathcal{A}^-_i) \big), \end{aligned} \label{eq:bounds-approximate}
    \end{align}
    where $c_i$ is either $c_i^{\mathrm{opt}}$ or $c_i^{\mathrm{ide}}$, $w(\mathcal{A}) = \overline{\mathcal{A}} - \underline{\mathcal{A}}$ is the width of an interval $\mathcal{A}$, and $K(\mathcal{A})$, for any $\mathcal{A} \in \mathbb{IR}^{n \times m}$, is given by
    \begin{align}
        K(\mathcal{A}) &= \begin{aligned}[t] \min \big( &\| 2 |S \mathcal{U}| + q +  2 | Q (\mathcal{B}_i + \mathcal{A} \mathcal{U} )| \|_2, \\ &\| 2|S \mathcal{U}| + q +  2| Q \mathcal{X}|\|_2 \big). \end{aligned} \nonumber
    \end{align}
\end{theorem}

We provide a proof for Theorem~\ref{thm:suboptimality-gap} in \myextended{} where the rate of change of the cost function values is exploited to characterize its variations over the over-approximating set~\eqref{eq:linear-over-approx}.


\begin{remark}
    The main term of the suboptimality bound in Theorem~\ref{thm:suboptimality-gap} is directly related to the width of the over-approximation of the reachable set at time $t_{i} + \Delta t$. Thus, as the over-approximation of the reachable set becomes tighter, the gap with the unknown optimal cost decreases. Specifically, it is straightforward to see that the more data and side information are used in the computation of $\mathcal{B}_i$, $\mathcal{A}^+_i$, and $\mathcal{A}^-_i$, the tighter the widths $w(\mathcal{B}_i)$, $w(\mathcal{A}^+_i)$, and $w(\mathcal{A}^-_i)$ become.
\end{remark}

\section{Comparison with Existing Approaches}
We compare \controlalgo{} with existing data-driven control algorithms \texttt{CGP-LCB}~\cite{krause2011contextual,gpyopt2016}, \texttt{C2Opt}~\cite{vinod2020convexified}, and \texttt{SINDYc}~\cite{kaiser2018sparse}. These algorithms also solve the one-step optimal control problem~\eqref{eq:one-step-optimal-control} approximately. \texttt{CGP-LCB} assumes that the unknown one-step cost $C(x_i,u_i) = c(x_i, u_i, x(t_{i+1}; x_i, u_i))$ to minimize is described by a Gaussian process,  while \texttt{C2Opt} assumes that $C$ has Lipschitz gradients. On the other hand, \texttt{SINDYc} uses the limited data to perform \emph{sparse identification} of the dynamics, which then permits to approximate a solution to \eqref{eq:one-step-optimal-control} via numerical solvers. The experiments on the unicycle system and a quadrotor demonstrate that \controlalgo{} outperforms these algorithms both on the computation time at each sampling time and the suboptimality of the control. See \texttt{\url{https://github.com/u-t-autonomous/ACC2021_DaTaReach_DaTaControl.git}} for reproducibility of the experiments demonstrated in this paper.
\begin{figure*}[t]
    \centering
    \input{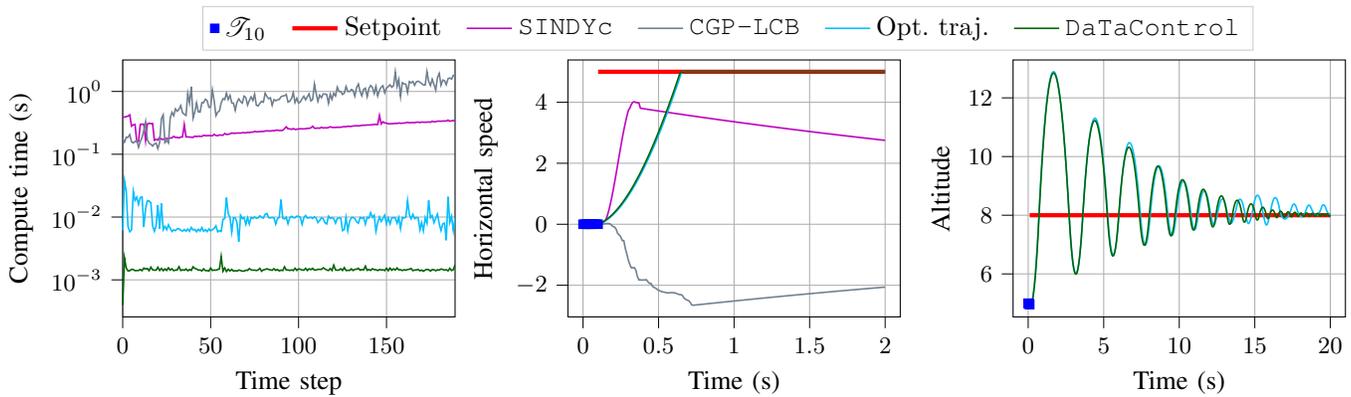}
    \vspace*{-5mm}
    \caption{\texttt{CGP-LCB} and \texttt{SINDYc} fail in the task of controlling the horizontal speed $v_x$ of the quadrotor whereas \controlalgo{} reaches both setpoints.}
    \vspace*{-3mm}
    \label{fig:quadrotor_traj_cost}
\end{figure*}

\subsection{Unicycle System}

We consider the problem of driving the unicycle of Example~\ref{ex:unicycle-system} to the origin. We encode this control objective using the one-step cost function $c(x_i,u_i,x_{i+1}) = 0.5 \|x_{i+1}\|^2_2$. We chose the time step size $\Delta t = 0.1s$ and generated a random initial trajectory $\mathscr{T}_{10}$ starting from the state $x_0= [-2,-2.5,\pi/2]$ and such that the unicycle goes away from the target. We applied \controlalgo{} with the side information discussed in Example~\ref{ex:unicycle-system}. We used the default parameters in \texttt{GPyOpt}~\cite{gpyopt2016} when implementing \texttt{CGP-LCB}. We chose the Lipschitz constant for the gradient $L=10$ and trade-off hyperparameter $\alpha=1/2$ for \texttt{C2Opt}~\cite{vinod2020convexified}. For \texttt{SINDYc}, we considered monomials (up to degree $6$), sines and cosines of the state, and the products of these functions with the velocity $v$ and the turning rate $\omega$  as the library functions. To perform sparse identification, we swept the regularization parameter~\cite{kaiser2018sparse} $\lambda\in\{10^{p}:p\in \mathbb{N}_{[-6,5]}\}$ and rounded-down the coefficients smaller than $10^{-3}$ to zero. We relaxed the target state to the $0.1$-sublevel set of the one-step cost function as the stopping criteria for the algorithms.

Figure~\ref{fig:unicycle_traj_cost} shows the trajectories and the evolution of the cost function for the different algorithms. \controlalgo{} performs significantly better than \texttt{GPyOpt}, \texttt{SINDYc}, and \texttt{C2Opt}. It reaches the origin in fewer time steps and significantly lower computation time. Figure~\ref{fig:unicycle_traj_cost} empirically shows that \controlalgo{} enables near-real-time control.

\subsection{Quadrotor System}
Consider a quadrotor with control-affine dynamics~\cite{decomposition-tomlin}
\begin{align*}
    &\dot{p}_x = v_x, &\dot{v}_x = -\frac{1}{m} C^v_D v_x-\frac{T_1}{m} sin \phi -\frac{T_2}{m} sin \phi, \\
    &\dot{p}_y = v_y, &\dot{v}_y = -\frac{1}{m}(m g+C^v_D v_y) + \frac{T_1}{m} cos \phi + \frac{T_2}{m} cos \phi, \\
    &\dot{\phi} = \omega,  &\dot{\omega} = -\frac{1}{2I_{yy}} C^\phi_D \omega -\frac{l}{2I_{yy}}T_1+\frac{l}{2I_{yy}} T_2,
\end{align*}
where the components of the state $x = [p_x,v_x,p_y,v_y,\phi,\omega]$ represent, respectively, the horizontal position, horizontal velocity, vertical position, vertical velocity, pitch angle, and pitch rate, and the components of the control $u = [T_1, T_2]$ represent the thrust exerted on either end of the quadrotor. We chose the constraint set $\mathcal{U} = [0,18.4] \times [0,18.4]$.  The constants of the dynamics are given by $C^v_D=0.25$, $C^\phi_D=0.02255$, $g=9.81$, $m=1.25$, $l=0.5$, and $I_{yy}=0.03$.

We assume that the dynamics of the quadrotor are unknown and consider the problems of controlling $v_x$ to a setpoint $v_x^{\mathrm{sp}} = 5$ and $p_y$ to a setpoint $p_y^{\mathrm{sp}} = 8$. We encode these control objectives, respectively, using the cost functions $c_1(x_i,u_i,x_{i+1}) = 0.5 ((x_{i+1})_2 - v_x^{\mathrm{sp}})^2$ and $c_2(x_i,u_i,x_{i+1}) = 0.5 ((x_{i+1})_3 - p_y^{\mathrm{sp}})^2$. We chose the time step size $\Delta t = 0.01s$ and generated a random initial trajectory $\mathscr{T}_{10}$ starting from the state $x_0= [0,0,5,0,0,0]$. We applied \controlalgo{} with the side information $\dot{p}_x = v_x$, $\dot{p}_y = v_y$, and $\dot{\phi} = \omega$. Such extra knowledge is obtained from elementary laws of physics. Furthermore, \controlalgo{} considered the loose Lipschitz bounds $L_{f_2}= L_{f_4} = 0.3$, $L_{f_6} = 0.9$, $L_{G_{6,1}}= L_{G_{6,2}} = 0.01$, $L_{G_{2,1}}= L_{G_{2,2}} = L_{G_{4,1}}= L_{G_{4,2}} = L_{G_{6,1}}= L_{G_{6,2}} = 0.9$, and uses the side information $G(x) = G(\phi)$ and $f(x) = f(v_x,v_y,w)$. We used the default parameters of \texttt{GPyOpt} when implementing \texttt{CGP-LCB}. For \texttt{SINDYc}, we considered monomials (up to degree $1$), sines and cosines of the state, and the products of these functions with the $T_1$ and $T_2$ as the library functions. We do not make a comparison with \texttt{C2Opt} due to the inability to compute the gradient of $C$.

Figure~\ref{fig:quadrotor_traj_cost} shows the near-optimality of \controlalgo{} while \texttt{CGP-LCB} and \texttt{SINDYc} fail to reach the setpoints. Besides, the figure demonstrates that \controlalgo{} can achieve near-real-time control of the vertical position and horizontal speed of the quadrotor. We justify the suboptimality of $\mathrm{Opt.\ traj.}$ by the fact that the one-step optimal control problem is highly nonlinear, which makes possible the synthesis of local optimum solutions by the numerical solvers.

\section{Conclusion} \label{sec:conclusion}

We develop two data-driven algorithms, \reachalgo{} and \controlalgo{}, for the on-the-fly over-approximation of the reachable set and constrained near-optimal control of systems with unknown dynamics. These algorithms are suitable for scenarios with severely limited data and can take advantage of various forms of side information on the underlying unknown dynamics. The numerical experiments empirically demonstrate the efficacy of the algorithms over existing approaches while suggesting that, in addition of being near-optimal, \controlalgo{} also achieves near-real-time control.

\bibliographystyle{IEEEtran}
\bibliography{IEEEabrv,ref}

\begin{appendices}

\section{}\label{appendix-A}
In this appendix, we provide the proofs for all lemmas and theorems presented in this paper.
\section*{Proof of Lemma~\ref{lem:contraction}}
    Given the knowledge that $f(x_i) \in \mathcal{F}_i$ and $G(x_i) \in \mathcal{G}_i$, we want to obtain tighter intervals $C_{\mathcal{F}_i} \subseteq \mathcal{F}_i$ and $C_{\mathcal{G}_i} \subseteq \mathcal{G}_i$ pruning out some values $f(x_i)$ and $G(x_i)$ from $\mathcal{F}_i$ and $\mathcal{G}_i$ that do not satisfy the constraint $\dot{x}_i = f(x_i) + G(x_i) u_i$. We first have that
    \begin{align*}
        f(x_i) = \dot{x}_i - G(x_i) u_i \in (\dot{x}_i - \mathcal{G}_i u_i) \cap \mathcal{F}_i = C_{\mathcal{F}_i}.
    \end{align*}
    Therefore, a similar reasoning using the tighter interval $C_{\mathcal{F}_i}$ provides that $G(x_i) u_i \in (\dot{x}_i - C_{\mathcal{F}_i}) \cap (\mathcal{G}_i u_i) = \mathcal{S}_0$. It is important to note that plugging back $\mathcal{S}_0$ instead of $\mathcal{G}_i u_i$ in the expression of $C_{\mathcal{F}_i}$ will not yield an interval tighter than $C_{\mathcal{F}_i}$. Therefore, $C_{\mathcal{F}_i}$ and $\mathcal{S}_0$ are optimal. Next, we focus on the term $G(x_i) u_i \in \mathcal{S}_0$. For all $k \in \mathbb{N}_{[1,n]}$, we have that
    \begin{align*}
        (G(x_i))_{k,1} (u_i)_1 &= (G(x_i) u_i)_k - \sum_{l > 1 } (G(x_i))_{k,l} (u_i)_l \\
        &\in \underbrace{\big( (\mathcal{S}_0)_k - \sum_{l>1} (\mathcal{G}_i)_{k,l} (u_i)_l \big) \cap \big( (\mathcal{G}_i)_{k,1} (u_i)_1\big)}_{(C_{\mathcal{G}_i})_{k,1} (u_i)_1},
    \end{align*}
    and we can deduce 
    \begin{align*}
        &\sum_{l > 1 } (G(x_i))_{k,l} (u_i)_l \\
            & \quad \quad \in \underbrace{\big( (\mathcal{S}_0)_k - (C_{\mathcal{G}_i})_{k,1} (u_i)_1 \big) \cap \big( \sum_{l>1} (\mathcal{G}_i)_{k,l} (u_i)_l \big)}_{(\mathcal{S}_1)_k}.
    \end{align*}
    Using the same argument as for the optimality of $\mathcal{S}_0$ and $C_{\mathcal{F}_i}$, we can say that $\mathcal{S}_1$ and $(C_{\mathcal{G}_i})_{k,1} (u_i)_1$ are optimal. Finally, we apply the previous step in a sequential manner for $p=2,\hdots,m$ to the equality $(G(x_i))_{k,p} (u_i)_p = \sum_{l > p-1 } (G(x_i))_{k,l} (u_i)_l - \sum_{l > p } (G(x_i))_{k,l} (u_i)_l$ in order to obtain optimal intervals $\mathcal{S}_p$ and $(C_{\mathcal{G}_i})_{k,p} (u_i)_p$. 
    The procedure described here is similar to the \texttt{HC4-Revise} algorithm \cite{benhamou-contractor1999}, where optimality was proven when each variable, in our case $f(x_i)$ and $G(x_i)$, appears only once in the definition of the constraint, in our case $\dot{x}_i = f(x_i) + G(x_i) u_i$.
\myqedblock

\section*{Proof of Lemma~\ref{lem:overapprox-f-G}}
    This is a direct result from combining the arithmetic of intervals and the definition of the Lipschitz bounds for each $f_k$ and $G_{k,l}$. Specifically, from the upper bound $L_{f_k}$ on the Lipschitz constant of $f_k$, we have that 
    \begin{align*}
        |(f(x))_k - (f(y))_k| \leq L_{f_k} \|x-y\|_2, \: \: \forall x,y \in \mathcal{X}.
    \end{align*}
    Hence, given a data $(x_i, C_{\mathcal{F}_i}, C_{\mathcal{G}_i}) \in \mathscr{E}_N$ and $x \in  \mathcal{X}$, we can write that $(f(x))_k \in (f(x_i))_k + L_{f_k} \|x-x_i\|_2 [-1,1]$, and therefore $(f(x))_k \in (C_{\mathcal{F}_i})_k + L_{f_k} \|x-x_i\|_2 [-1,1]$ since $f(x_i) \in C_{\mathcal{F}_i}$. The previous belonging relation is valid for every data $(x_i, C_{\mathcal{F}_i}, C_{\mathcal{G}_i)} \in \mathscr{E}_N$ and as a result $(f(x))_k \in (\boldsymbol{f}(x))_k$. The same reasoning applied to $L_{G_{k,l}}$ enables to show that $(G(x))_{k,l} \in (\boldsymbol{G}(x))_{k,l}$.
\myqedblock

\section*{Proof of Theorem~\ref{thm:diff-inclusion}}
    This result is a direct consequence of Lemma~\ref{lem:contraction} and Lemma~\ref{lem:overapprox-f-G}. First, we show that for all $(x_i, C_{\mathcal{F}_i}, C_{\mathcal{G}_i}) \in \mathscr{E}_N$ given by Algorithm~\ref{algo:overapprox-datapoints}, we have $f(x_i) \in C_{\mathcal{F}_i}$ and $G(x_i) \in C_{\mathcal{G}_i}$. Specifically, as a consequence of line~\ref{alg:update-ei} of Algorithm~\ref{algo:overapprox-datapoints} and Lemma~\ref{lem:overapprox-f-G}, we have that $f(x_i) \in \mathcal{F}_i$ and $G(x_i) \in \mathcal{G}_i$. Hence, by line~\ref{alg:contraction-f-G} and Lemma~\ref{lem:contraction}, we immediately have that $f(x_i) \in C_{\mathcal{F}_i}$ and $G(x_i) \in C_{\mathcal{G}_i}$. As a consequence of lines~\ref{alg:begin-init-e}--\ref{alg:end-init-e}, $\mathscr{E}_N$ can be used in Lemma~\ref{lem:overapprox-f-G} to conclude that $f(x) \in \boldsymbol{f}(x)$ and $G(x) \in \boldsymbol{G}(x)$ for all $x \in \mathcal{X}$. Therefore, using interval arithmetic, we have $\dot{x} = f(x) + G(x) u \in \boldsymbol{f}(x) + \boldsymbol{G}(x) u$. The lines~\ref{alg:while-begin}--\ref{alg:while-end} enable to obtain tighter sets $C_{\mathcal{F}_i}$ and $C_{\mathcal{G}_i}$ by combining the Lipschitz continuity, the trajectory, and the values $C_{\mathcal{F}_i}$ and $C_{\mathcal{G}_i}$ obtained during previous iterations. Hence, $\mathscr{E}_N$ is optimal given the trajectory and the Lipschitz continuity since we have the optimality of $C_{\mathcal{F}_i}$ and $C_{\mathcal{G}_i}$ by Lemma~\ref{lem:contraction}.
\myqedblock

\section*{Proof of Theorem~\ref{thm:over-approximation-state}}
    This proof applies the interval Taylor-based method of Section~\ref{sec:prelem-taylor-method} on the dynamics described by $\dot{x}\in \boldsymbol{f}(x) + \boldsymbol{G}(x) u$. Clearly, $\boldsymbol{h}(\mathcal{Z},\mathcal{W}) = \boldsymbol{f}(\mathcal{Z}) + \boldsymbol{G}(\mathcal{Z}) \mathcal{W}$ is an interval extension of $h(x,u) = f(x) + G(x) u$.
    Hence, by simple inclusion between $h$ and $\boldsymbol{h}$, the set $\mathcal{S}_i$ solution of~\eqref{eq:fix-point-rough} is an a priori rough enclosure as it satisfies the fixed-point equation~\eqref{eq:rough-enclosure-approx}. We use a Taylor expansion~\eqref{eq:taylor-expansion} of order $D=2$ to obtain
    \begin{align}
        \mathcal{R}^+_{i+1} &\subseteq  \mathcal{R}^+_i + \Delta t \Big( \boldsymbol{h}(\mathcal{R}^+_i,\boldsymbol{v}(t_i)) \Big) + \frac{\Delta t^2}{2} \Big(  \boldsymbol{\frac{\partial h }{\partial x}}  \boldsymbol{h}  \Big) (\mathcal{S}_i, \mathcal{V}_i)
                                              \nonumber \\
                                              & \quad \quad \quad + \frac{\Delta t^2}{2} \Big( \boldsymbol{\frac{\partial h }{\partial u}} \Big) (\mathcal{S}_i, \mathcal{V}_i) \mathcal{V}_i^{(1)}, \label{eq:taylor-2-order}
    \end{align}
    where $\boldsymbol{\frac{\partial h }{\partial x}}$ and $\boldsymbol{\frac{\partial h }{\partial u}}$ are interval extensions of $\frac{\partial h }{\partial x}$ and $\frac{\partial h }{\partial u}$, respectively. We have that $\boldsymbol{G}$ is an interval extension of $\frac{\partial h }{\partial u}$ as
    \begin{align} \label{eq:proof-partial-der-u-extension}
        \frac{\partial h }{\partial u} (x,u) = G(x) \in \boldsymbol{G}(x) \Longrightarrow \boldsymbol{\frac{\partial h }{\partial u}}  = \boldsymbol{G}. 
    \end{align}
    Furthermore, for all $k,p \in \mathbb{N}_{[1,n]}$, $x \in \mathcal{X}$ and $u \in  \mathcal{U}$, we have 
    \begin{align*}
        \frac{\partial h_k}{\partial x_p}(x,u) = \frac{\partial f_k}{\partial x_p}(x,u) + \sum_{l=1}^m \frac{\partial G_{k,l}}{ \partial x_p}(x,u) (u)_l.
    \end{align*}
    Thus, a consequence of \textit{Rademacher's theorem}~\cite[Theorem $3.1.6$]{Federer:228981} enables to write that
    \begin{align*}
        \frac{\partial f_k}{\partial x_p}  \in  L_{f_k} [-1, 1] \text{ and }
        \frac{\partial G_{k,l}}{ \partial x_p}  \in L_{G_{k,l}}[-1,1].
    \end{align*}
    Therefore, we have that $\boldsymbol{\frac{\partial h }{\partial x}}(\mathcal{S}_i, \mathcal{V}_i) = \mathcal{J}_f + \mathcal{J}_G \mathcal{V}_i$.
    Finally, merging $\boldsymbol{\frac{\partial h }{\partial x}}$ and $\boldsymbol{\frac{\partial h }{\partial u}}$ into~\eqref{eq:taylor-2-order} provides the over-approximating set~\eqref{eq:over-approx-next-state}.
\myqedblock

\section*{Proof of Theorem~\ref{thm:linear-expr-O}}
    We  apply Theorem~\ref{thm:over-approximation-state} with $\mathbb{V} = \{u\}$. Specifically, since $u$ is such that  $\dot{u}(t) = 0$ for all $t \in [t_i , t_{i}+\Delta t]$,  we have that $\boldsymbol{v}(t_i)=u_i$, $\boldsymbol{v}([t_i,t_i+\Delta t]) = u_i$, and $\boldsymbol{v}^{(1)}([t_i,t_i+\Delta t]) = 0$. Applying Theorem~\ref{thm:over-approximation-state}, $\mathcal{R}^+_{i+1}$ as a function of $u_i$ is therefore reduced to a term independent of $u_i$, a term linear in $u_i$, and a term quadratic in $u_i$.
    
    Specifically, it is immediate from~\eqref{eq:over-approx-next-state} that the term independent of $u_i$ is given by $\mathcal{B}_i$~\eqref{eq:linear-over}. The term linear in $u_i$ is
    \begin{align}
        & \big( \boldsymbol{G} (\mathcal{R}^+_i) \Delta t \big) u_i  +  \big( \mathcal{J}_G u_i \boldsymbol{f} (\mathcal{S}_j) + \mathcal{J}_f \boldsymbol{G}(\mathcal{S}_j) u_i \big) \frac{\Delta t^2}{2}. \label{eq:full-linear-term}
    \end{align}
    The $k$-th component of $\mathcal{J}_G u_i \boldsymbol{f} (\mathcal{S}_i)$ is given by
    \begin{align}
        \sum_{p=1}^n {(\mathcal{J}_G u_i)}_{k,p} (\boldsymbol{f} {(\mathcal{S}_i))}_p &= \sum_{p=1}^n \big( \sum_{l=1}^m {(\mathcal{J}_G)}_{k,l,p} {(u_i)}_l \big) (\boldsymbol{f} {(\mathcal{S}_i))}_p \nonumber \\
        &= \sum_{l=1}^m \big( \underbrace{\sum_{p=1}^n  {(\mathcal{J}_G)}_{k,l,p} {(\boldsymbol{f} (\mathcal{S}_i))}_p}_{{(\mathcal{J}_G^\textrm{T} \boldsymbol{f}(\mathcal{S}_i))}_{k,l}} \big) {(u_i)}_l \nonumber \\
        &= {(\mathcal{J}_G^\textrm{T} \boldsymbol{f}(\mathcal{S}_i) u_i)}_k. \label{eq:linear-reordering}
    \end{align}
    The quadratic term $0.5 \Delta t^2 \mathcal{J}_G u_i \boldsymbol{G} (\mathcal{S}_i) u_i$ in $u_i$ satisfies
    \begin{align}
         \mathcal{J}_G u_i \boldsymbol{G} (\mathcal{S}_i) u_i &\subseteq  \mathcal{J}_G \mathcal{U} \boldsymbol{G}(\mathcal{S}_i) u_i, \label{eq:a1-formula}\\
         \mathcal{J}_G u_i \boldsymbol{G} (\mathcal{S}_i) u_i &\subseteq  \mathcal{J}_G u_i \boldsymbol{G}(\mathcal{S}_i) \mathcal{U} =  \mathcal{J}^{\textrm{T}}_G \boldsymbol{G}(\mathcal{S}_i) \mathcal{U} u_i. \label{a2-formula}
    \end{align}
    Henceforth, by combining~\eqref{eq:full-linear-term},~\eqref{eq:linear-reordering}, and~\eqref{eq:a1-formula}, the linear and quadratic terms in $u_i$ of $\mathcal{R}^+_{i+1}$ can be over-approximated by the linear term  $\mathcal{A}^+_i u_i$ with $\mathcal{A}^+_i$ given by~\eqref{eq:linear-over}. Similarly, using~\eqref{eq:full-linear-term},~\eqref{eq:linear-reordering}, and~\eqref{a2-formula} helps to prove that the linear and quadratic terms in $u_i$ can be also over-approximated by $\mathcal{A}^-_i u_i$ with $\mathcal{A}^-_i$ given by~\eqref{eq:linear-over}. Hence, we deduce the intersection in~\eqref{eq:linear-over-approx}. 
\myqedblock

\section*{Proof of Theorem~\ref{thm:suboptimality-gap}}
        For any $x^+_{i+1} = b_i + A^+_i u_i  \in \mathcal{B}_i + \mathcal{A}^+_i u_i$ and $\hat{x}^+_{i+1} = \hat{b}_i + \hat{A}_i^+ u_i \in \mathcal{B}_i + \mathcal{A}^+_i u_i$ with $u_i \in \mathcal{U}$, we have that 
    \begin{align}
        & \quad |c(\cdot, u_i , x^+_{i+1}) - c(\cdot,u_i,\hat{x}^+_{i+1})| \nonumber \\
        &= \big| \begin{bmatrix} x^+_{i+1} \\ u_i \end{bmatrix}^{\mathrm{T}} \begin{bmatrix} Q & S \\ S^{\mathrm{T}}  & R\end{bmatrix} \begin{bmatrix} x^+_{i+1} \\ u_i \end{bmatrix} 
        - \begin{bmatrix} \hat{x}^+_{i+1} \\ u_i \end{bmatrix}^{\mathrm{T}} \begin{bmatrix} Q & S \\ S^{\mathrm{T}}  & R\end{bmatrix} \begin{bmatrix} \hat{x}^+_{i+1} \\ u_i \end{bmatrix} \nonumber \\
        & \quad \quad  + \begin{bmatrix} q \\ r\end{bmatrix}^{\mathrm{T}} \begin{bmatrix}  x^+_{i+1} - \hat{x}^+_{i+1} \\ 0 \end{bmatrix} \big| \nonumber \\
        &=  \big| \big( x^+_{i+1} - \hat{x}^+_{i+1} \big)^{\mathrm{T}}  \big( Q \big( x^+_{i+1} + \hat{x}^+_{i+1} \big) +  2 S u_i + q \big) \big| \nonumber \\
        &\leq \big(\| (b_i - \hat{b}_i) + (A^+_i - \hat{A}^+_i) u_i \|_2 \big) \nonumber \\ 
        &\quad \quad \big( \|  Q \big( (b_i + \hat{b}_i) + (A^+_i + \hat{A}^+_i) u_i \big) + 2 S u_i + q \|_2 \big) \label{eq:diff-cost}.
    \end{align}
    By definition of the width  $w$, $|b_i - \hat{b}_i| \leq w(\mathcal{B}_i)$ and $|A^+_i - \hat{A}^+_i| \leq w(\mathcal{A}^+_i)$. Hence, we can deduce that
    \begin{align}
        \| (b_i - \hat{b}_i) + (A^+_i - \hat{A}^+_i) u_i \|_2 \leq \| w(\mathcal{B}_i) + w(\mathcal{A}^+_i)|\mathcal{U}| \|_2. \label{eq:first-term}
    \end{align}
    Furthermore, interval arithmetic provides that
    \begin{align*}
        (b_i + \hat{b}_i) + (A^+_i + \hat{A}^+_i) u_i \in \big( 2\mathcal{B}_i + 2\mathcal{A}^+_i u_i \big)  \cap  2\mathcal{X}.
    \end{align*}
    Hence, we have
    \begin{align}
        &\begin{aligned}[t] \|& 2 S u_i + q + Q \big( (b_i + \hat{b}_i) + (A^+_i + \hat{A}^+_i) u_i \big) \|_2  \\
        &\quad \leq \| 2 |S \mathcal{U}| + q +  2 | Q (\mathcal{B}_i + \mathcal{A}^+_i \mathcal{U} )| \|_2, 
        \end{aligned} \label{eq:second-term-unbound} \\
        &\begin{aligned}[t]
        &\| 2 S u_i + q + Q \big( (b_i + \hat{b}_i) + (A^+_i + \hat{A}^+_i) u_i \big) \|_2 \\
        &\quad \leq  \| 2|S \mathcal{U}| + q +  2| Q \mathcal{X}|\|_2. \label{eq:second-term-bound}
        \end{aligned}
    \end{align}
    Therefore, combining~\eqref{eq:second-term-bound} and~\eqref{eq:second-term-unbound}, we have that
    \begin{align}
        &\| 2 S u_i + q + Q \big( (b_i + \hat{b}_i) + (A^+_i + \hat{A}^+_i) u_i \big) \|_2 \leq K(\mathcal{A}^+_i). \label{eq:m_plus_bound}
    \end{align}
    As a consequence of~\eqref{eq:diff-cost},~\eqref{eq:first-term}, and~\eqref{eq:m_plus_bound}, we can write 
    \begin{align}
        &|c(\cdot, u_i , x^+_{i+1}) - c(\cdot,u_i,\hat{x}^+_{i+1})| \nonumber\\  \leq & \| w(\mathcal{B}_i) + w(\mathcal{A}^+_i)|\mathcal{U}| \|_2 K(\mathcal{A}^+_i). \label{eq:bound-diff-cost}
    \end{align}
    
    Let $u^*_i$ be the optimal solution of the one-step optimal control problem~\eqref{eq:one-step-optimal-control} and $x_{i+1}(u^*_i) \in \mathcal{B}_i + \mathcal{A}^+_i u^*_i$ be the corresponding unknown next state. By optimality of $u^*_i$, we have that $c(x_i, u^*_i, x_{i+1}(u^*_i)) \leq c(x_i, u_i, x_{i+1}(u_i))$ for all $u_i \in \mathcal{U}$. If $\hat{u}_i$ is the optimal solution of the optimistic control problem~\eqref{eq:optimistic-control-problem} and $\hat{x}_{i+1}(\hat{u}_i) \in \mathcal{B}_i + \mathcal{A}^+_i u^*_i$ is the known next state for which the optimum of~\eqref{eq:optimistic-control-problem} is attained, then
    \begin{align*}
      |c_i^{\star} - c_i^{\mathrm{opt}}| &=  | c(x_i, u^*_i, x_{i+1}(u^*_i)) - c(x_i, \hat{u}_i, \hat{x}_{i+1}(\hat{u}_i)) |\\
                            &\leq  | c(x_i, \hat{u}_i, x_{i+1}(\hat{u}_i)) - c(x_i, \hat{u}_i, \hat{x}_{i+1}(\hat{u}_i)) | \\
                            &\leq \| w(\mathcal{B}_i) + w(\mathcal{A}^+_i)|\mathcal{U}| \|_2 K(\mathcal{A}^+_i).
    \end{align*}
    Similarly, using $\mathcal{A}^-_i$ instead of $\mathcal{A}^+_i$, we can prove that $|c_i^{\star} - c_i^{\mathrm{opt}}| \leq \| w(\mathcal{B}_i) + w(\mathcal{A}^-_i)|\mathcal{U}| \|_2 K(\mathcal{A}^-_i)$. Hence we obtain the suboptimality bounds~\eqref{eq:bounds-approximate} for the optimistic control problem.
    
    Finally, let $\hat{u}_i$ be the optimal solution of the idealistic control problem~\eqref{eq:idealistic-control-problem} and $\hat{x}_{i+1}(\hat{u}_i) = b^{\mathrm{ide}}_i + A^{\mathrm{ide}}_i \hat{u}_i$ be the next state for which the optimum is attained. We have that
    \begin{align*}
        |c_i^{\mathrm{ide}} - c_i^{\star}| &= \begin{cases} c_i^{\mathrm{ide}} - c(x_i, u^{\star}_i, x_{i+1}(u^*_i)),& \text{if } c_i^{\mathrm{ide}} \geq c_i^{\star} \\ c_i^{\star} - c(x_i, \hat{u}_i,\hat{x}_{i+1}(\hat{u}_i) ),& \text{otherwise} \end{cases} \\
        &\leq \begin{cases} c(x_i, u^{\star}_i, \hat{x}_{i+1}(u^{\star}_i)) - c(x_i, u^{\star}_i, x_{i+1}(u^*_i)),\\ \quad \text{if } c_i^{\mathrm{ide}} \geq c_i^{\star} \\ c(x_i, \hat{u}_i, x_{i+1}(\hat{u}_i))  - c(x_i, \hat{u}_i,\hat{x}_{i+1}(\hat{u}_i) ), \\ \quad \text{otherwise} \end{cases} \\
        &\leq \begin{aligned}[t] \max \big( &\| w(\mathcal{B}_i) + w(\mathcal{A}^+_i)|\mathcal{U}| \|_2 K (\mathcal{A}^+_i) , \\ 
        &\| w(\mathcal{B}_i) + w(\mathcal{A}^-_i)|\mathcal{U}| \|_2  K (\mathcal{A}^-_i) \big), \end{aligned}
    \end{align*}
    where the second inequality follows from the definition of optimality of~\eqref{eq:one-step-optimal-control} and~\eqref{eq:idealistic-control-problem}, and the last inequality follows similarly to the inequalities obtained for the optimistic suboptimality bounds. 
\myqedblock

\end{appendices}

\end{document}

%% file: derpx_unicycle_evolution.tex
\begin{tikzpicture}

\definecolor{color0}{rgb}{1,0,0}

\begin{axis}[
width=9cm,
height=4.1cm,
legend cell align={left},
legend columns=4,
legend style={fill opacity=1.0, draw opacity=1, text opacity=1, at={(-0.0,1.)}, anchor=south west, draw=white!80!black, /tikz/every even column/.append style={column sep=0.2cm}},
tick align=outside,
tick pos=left,
x grid style={white!69.0196078431373!black},
xlabel={Time (s)},
xmajorgrids,
xmin=-0.2, xmax=4.2,
xtick style={color=black},
y grid style={white!69.0196078431373!black},
ylabel={\(\displaystyle \dot{p}_x\)},
ymajorgrids,
ymin=-0.631342019899496, ymax=0.623492451935764,
ytick style={color=black},
ytick={-0.8,-0.6,-0.4,-0.2,0,0.2,0.4,0.6,0.8},
yticklabels={-0.8,-0.6,-0.4,-0.2,0.0,0.2,0.4,0.6,0.8}
]
\path [draw=green!39.2156862745098!black, fill=green!50.1960784313725!black, opacity=0.8]
(axis cs:0,0)
--(axis cs:0,0)
--(axis cs:0.05,-3.29476322889599e-05)
--(axis cs:0.1,0.00429913825104174)
--(axis cs:0.15,0.00429913811420159)
--(axis cs:0.2,0.0153811878091318)
--(axis cs:0.25,0.283923072753867)
--(axis cs:0.3,0.563555025605114)
--(axis cs:0.35,0.33323697081861)
--(axis cs:0.4,0.000781281643900524)
--(axis cs:0.45,0.00225710809472055)
--(axis cs:0.5,0.504049841506768)
--(axis cs:0.55,0.153695287997693)
--(axis cs:0.6,-0.171928272352215)
--(axis cs:0.65,-0.57430408936153)
--(axis cs:0.7,-0.437893031644485)
--(axis cs:0.75,-0.244051715671409)
--(axis cs:0.8,-0.0196073355560675)
--(axis cs:0.85,-0.0196073372397245)
--(axis cs:0.9,-0.0153269271120648)
--(axis cs:0.95,-0.210890217737506)
--(axis cs:1,-0.533545876818212)
--(axis cs:1.05,-0.452848764092935)
--(axis cs:1.1,-0.275550372353579)
--(axis cs:1.15,-0.070466988974358)
--(axis cs:1.2,0)
--(axis cs:1.25,-0.00117269700830653)
--(axis cs:1.3,0)
--(axis cs:1.35,-4.72127648087906e-11)
--(axis cs:1.4,0)
--(axis cs:1.45,-0.000367242411932298)
--(axis cs:1.5,-0.260357547595411)
--(axis cs:1.55,-0.297974262566514)
--(axis cs:1.6,-0.324990307877348)
--(axis cs:1.65,-0.336881462561773)
--(axis cs:1.7,-0.332585526343442)
--(axis cs:1.75,-0.312486242268442)
--(axis cs:1.8,-0.278379018140866)
--(axis cs:1.85,-0.237495095797155)
--(axis cs:1.9,-0.20886053611327)
--(axis cs:1.95,-0.15357038339725)
--(axis cs:2,-0.101022414159646)
--(axis cs:2.05,-0.0562177141749289)
--(axis cs:2.1,-0.0224373786485097)
--(axis cs:2.15,-0.00475201289190693)
--(axis cs:2.2,-0.00130463193267606)
--(axis cs:2.25,-0.0124031804216101)
--(axis cs:2.3,-0.0407174672211829)
--(axis cs:2.35,-0.0813555670581123)
--(axis cs:2.4,-0.131061468818288)
--(axis cs:2.45,-0.185863905737735)
--(axis cs:2.5,-0.240355702004753)
--(axis cs:2.55,-0.260753099010521)
--(axis cs:2.6,-0.299852868788224)
--(axis cs:2.65,-0.326073202069873)
--(axis cs:2.7,-0.337071914325107)
--(axis cs:2.75,-0.331866522233487)
--(axis cs:2.8,-0.310922007298465)
--(axis cs:2.85,-0.276109282720037)
--(axis cs:2.9,-0.23472261145041)
--(axis cs:2.95,-0.205777405297079)
--(axis cs:3,-0.150502879106902)
--(axis cs:3.05,-0.0982946057112389)
--(axis cs:3.1,-0.0540180001200506)
--(axis cs:3.15,-0.0209351821424822)
--(axis cs:3.2,-0.00412188193273407)
--(axis cs:3.25,-0.00160285605400814)
--(axis cs:3.3,-0.0136031201874076)
--(axis cs:3.35,-0.0427119338476589)
--(axis cs:3.4,-0.0839362125990465)
--(axis cs:3.45,-0.134055454473979)
--(axis cs:3.5,-0.188973601610055)
--(axis cs:3.55,-0.239925875066802)
--(axis cs:3.6,-0.263229893718937)
--(axis cs:3.65,-0.301690711120046)
--(axis cs:3.7,-0.327107920552478)
--(axis cs:3.75,-0.337211078941055)
--(axis cs:3.8,-0.331097702856003)
--(axis cs:3.85,-0.309313881751547)
--(axis cs:3.9,-0.273805497398476)
--(axis cs:3.95,-0.231928954718764)
--(axis cs:4,-0.202687761203413)
--(axis cs:4,-0.176150526384413)
--(axis cs:4,-0.176150526384413)
--(axis cs:3.95,-0.225053187742131)
--(axis cs:3.9,-0.24266709157919)
--(axis cs:3.85,-0.20715870722612)
--(axis cs:3.8,-0.185374886121664)
--(axis cs:3.75,-0.179261510036612)
--(axis cs:3.7,-0.189364668425188)
--(axis cs:3.65,-0.21478187785762)
--(axis cs:3.6,-0.253242695258729)
--(axis cs:3.55,-0.216152364871679)
--(axis cs:3.5,-0.16268346876571)
--(axis cs:3.45,-0.108754837794248)
--(axis cs:3.4,-0.0586489966195929)
--(axis cs:3.35,-0.0182803703449953)
--(axis cs:3.3,0.000424129234100847)
--(axis cs:3.25,0.000254874181220586)
--(axis cs:3.2,-0.00221876384383461)
--(axis cs:3.15,0.0031040074555836)
--(axis cs:3.1,-0.0295634872819181)
--(axis cs:3.05,-0.0732708464861374)
--(axis cs:3,-0.124905912433785)
--(axis cs:2.95,-0.17918450121513)
--(axis cs:2.9,-0.227898104230321)
--(axis cs:2.85,-0.24036330625763)
--(axis cs:2.8,-0.205550581679201)
--(axis cs:2.75,-0.184606066744179)
--(axis cs:2.7,-0.17940067465256)
--(axis cs:2.65,-0.190399386907794)
--(axis cs:2.6,-0.216619720189443)
--(axis cs:2.55,-0.255719489967145)
--(axis cs:2.5,-0.213204745433731)
--(axis cs:2.45,-0.159629803449648)
--(axis cs:2.4,-0.105814797826047)
--(axis cs:2.35,-0.0560381641285756)
--(axis cs:2.3,-0.0163218400541314)
--(axis cs:2.25,-0.000775810531696588)
--(axis cs:2.2,0.000547724894961097)
--(axis cs:2.15,-0.00283754109203138)
--(axis cs:2.1,0.00162887755326832)
--(axis cs:2.05,-0.0317632013367965)
--(axis cs:2,-0.0760487064657084)
--(axis cs:1.95,-0.127918146376559)
--(axis cs:1.9,-0.182212080124723)
--(axis cs:1.85,-0.230721459849483)
--(axis cs:1.8,-0.2380935708368)
--(axis cs:1.75,-0.203986346709224)
--(axis cs:1.7,-0.183887062634224)
--(axis cs:1.65,-0.179591126415893)
--(axis cs:1.6,-0.191482281100319)
--(axis cs:1.55,-0.218498326411152)
--(axis cs:1.5,-0.256115041382256)
--(axis cs:1.45,0.000367242411932298)
--(axis cs:1.4,0)
--(axis cs:1.35,4.72127648087906e-11)
--(axis cs:1.3,0)
--(axis cs:1.25,0.00117269700830653)
--(axis cs:1.2,0)
--(axis cs:1.15,-0.0628961240296168)
--(axis cs:1.1,-0.26841272543351)
--(axis cs:1.05,-0.418653312120002)
--(axis cs:1,-0.52225958533107)
--(axis cs:0.95,-0.187796713492809)
--(axis cs:0.9,-0.0153020757190835)
--(axis cs:0.85,-0.0195790371412248)
--(axis cs:0.8,-0.0195790388248818)
--(axis cs:0.75,-0.218438905581865)
--(axis cs:0.7,-0.426194302469795)
--(axis cs:0.65,-0.524364192156325)
--(axis cs:0.6,-0.170066829078329)
--(axis cs:0.55,0.209082005634469)
--(axis cs:0.5,0.508194651933879)
--(axis cs:0.45,0.00370127026814604)
--(axis cs:0.4,0.00173939196242205)
--(axis cs:0.35,0.38883684315089)
--(axis cs:0.3,0.566454521397797)
--(axis cs:0.25,0.342595827160605)
--(axis cs:0.2,0.0153812209506172)
--(axis cs:0.15,0.00429917120122118)
--(axis cs:0.1,0.00429917106438103)
--(axis cs:0.05,3.29476322889599e-05)
--(axis cs:0,0)
--cycle;
\addlegendimage{area legend, draw=green!39.2156862745098!black, fill=green!50.1960784313725!black, opacity=0.8}
\addlegendentry{$\boldsymbol{f}(x) + \boldsymbol{G}(x) u$}

\addplot [color0]
table {%
0 0
0.05 0
0.1 0.00429915465771138
0.15 0.00429915453502348
0.2 0.0153812043798745
0.25 0.311600267073578
0.3 0.565004773501456
0.35 0.35268282110928
0.4 0.00126033680316128
0.45 0.00293734967588392
0.5 0.506122246720323
0.55 0.191049667799939
0.6 -0.170997550091243
0.65 -0.532861200707638
0.7 -0.432043666433111
0.75 -0.227003497941741
0.8 -0.0195931865664455
0.85 -0.019593188105405
0.9 -0.0153145014155741
0.95 -0.196293028438715
1 -0.527902731074641
1.05 -0.436619801132832
1.1 -0.271981548893544
1.15 -0.0652788775807478
1.2 0
1.25 0
1.3 0
1.35 0
1.4 0
1.45 0
1.5 -0.257309194680127
1.55 -0.292350700393367
1.6 -0.315961000029224
1.65 -0.326292448867049
1.7 -0.322564406177796
1.75 -0.305056495315404
1.8 -0.275112504400419
1.85 -0.235136984488259
1.9 -0.188514714303708
1.95 -0.139383267645088
2 -0.0922447932043304
2.05 -0.0514790643283786
2.1 -0.0208653798159846
2.15 -0.00320783156524225
2.2 -0.000102088454699857
2.25 -0.0118268001702835
2.3 -0.0373267691220784
2.35 -0.0742801237391835
2.4 -0.119275631976202
2.45 -0.168129489353442
2.5 -0.216329219806849
2.55 -0.259530634733473
2.6 -0.293998459987244
2.65 -0.316903430994444
2.7 -0.326457607301282
2.75 -0.321939958471794
2.8 -0.30368950776509
2.85 -0.273109936034145
2.9 -0.232663978721872
2.95 -0.185786196552373
3 -0.136646197562956
3.05 -0.0897526004740712
3.1 -0.0494635504120995
3.15 -0.0195123255636365
3.2 -0.00264014845084154
3.25 -0.000370758825961421
3.3 -0.0129077442592845
3.35 -0.0391222725031784
3.4 -0.0766255195684724
3.45 -0.12195322379167
3.5 -0.170890473790887
3.55 -0.218921085729309
3.6 -0.261724391295767
3.65 -0.295609619787122
3.7 -0.317803642957587
3.75 -0.326578283579031
3.8 -0.321272091745929
3.85 -0.302283512191388
3.9 -0.271076114764845
3.95 -0.230170565057614
4 -0.18305047678589
};
\addlegendentry{$f(x) + G(x) u$}
\addplot [semithick, blue, mark=square*, mark size=1.5, mark options={solid}, only marks]
table {%
0 0
0.1 0.00429915465771138
0.2 0.0153812043798745
0.3 0.565004773501456
0.4 0.00126033680316128
0.5 0.506122246720323
0.6 -0.170997550091243
0.7 -0.432043666433111
0.8 -0.0195931865664455
0.9 -0.0153145014155741
1 -0.527902731074641
1.1 -0.271981548893544
1.2 0
1.3 0
1.4 0
};
\addlegendentry{$\mathscr{T}_{15}$}
\end{axis}

\end{tikzpicture}